\documentclass[manuscript,screen]{acmart}

\AtBeginDocument{%
  \providecommand\BibTeX{{%
    \normalfont B\kern-0.5em{\scshape i\kern-0.25em b}\kern-0.8em\TeX}}}

\setcopyright{acmlicensed}
\acmJournal{CSUR}
\acmYear{2023} \acmVolume{1} \acmNumber{1} \acmArticle{1} \acmMonth{1} \acmPrice{15.00}\acmDOI{10.1145/3626516}

\usepackage{amsmath}
\usepackage{graphicx}
\usepackage{mathtools}
\usepackage{balance}
\usepackage{graphics} 
\usepackage[T1]{fontenc}
\usepackage{txfonts}
\usepackage{mathptmx}
\usepackage{color, colortbl}
\usepackage{booktabs}
\usepackage{textcomp}
\usepackage{float}
\usepackage{enumitem}
\usepackage{caption}
\usepackage{multirow}
\usepackage{longtable}
\usepackage{arydshln}
\usepackage{hyperref}

\definecolor{LightCyan}{rgb}{0.93,0.95,1.0}

\begin{document}
\title{Co-Located Human-Human Interaction Analysis using Nonverbal Cues: A Survey}

\author{Cigdem Beyan}
\email{cigdem.beyan@unibg.it}
\affiliation{%
  \institution{
  Dept. of Management, Information and Production Engineering, University of Bergamo}
  \streetaddress{Via A. Einstein 2}
  \city{Dalmine}
  \postcode{24044}
  \country{Italy}
  }
 
\author{Alessandro Vinciarelli}
\email{Alessandro.Vinciarelli@glasgow.ac.uk}
\affiliation{
  \institution{
  School of Computing Science, University of Glasgow}
  \city{Glasgow}
  \postcode{G12 8QQ}
  \country{UK}}

\author{Alessio {Del Bue}}
\email{Alessio.DelBue@iit.it}
  \affiliation{%
  \institution{
  Pattern Analysis and Computer Vision, Istituto Italiano di Tecnologia}
  \streetaddress{Enrico Melen 83}
  \city{Genoa}
  \postcode{16152}
  \country{Italy}}

\renewcommand{\shortauthors}{Beyan et al.}

\text{© [Beyan et al.] [2023]. This is the author's version of the work.  It is posted here for your personal use. Not for redistribution.}
\text{
The definitive version was published in ACM Computing Surveys}
\text{URL: https://dl.acm.org/journal/csur} 
\text{DOI: https://doi.org/10.1145/3626516}
\text{}
\text{}
\text{}

\begin{abstract}
Automated co-located human-human interaction analysis has been addressed by the use of nonverbal communication as measurable evidence of social and psychological phenomena. We survey the computing studies (since 2010) detecting phenomena related to social traits (e.g., leadership, dominance, personality traits), social roles/relations, and interaction dynamics (e.g., group cohesion, engagement, rapport). Our target is to identify the nonverbal cues and computational methodologies resulting in effective performance. This survey differs from its counterparts by involving the widest spectrum of social phenomena and interaction settings (free-standing conversations, meetings, dyads, and crowds). We also present a comprehensive summary of the related datasets and outline future research directions which are regarding the implementation of artificial intelligence, dataset curation, and privacy-preserving interaction analysis. Some major observations are: the most often used nonverbal cue, computational method, interaction environment, and sensing approach are speaking activity, support vector machines, and meetings composed of 3-4 persons equipped with microphones and cameras, respectively; multimodal features are prominently performing better; deep learning architectures showed improved performance in overall, but there exist many phenomena whose detection has never been implemented through deep models. We also identified several limitations such as the lack of scalable benchmarks, annotation reliability tests, cross-dataset experiments, and explainability analysis.
\end{abstract}



\maketitle

\section{Introduction}
\label{intro}
Human Behavior Understanding (HBU) is an important topic in research domains; video surveillance, social robotics, detection of mental health issues, human-computer interaction, and many others. Until now, HBU has mostly been performed in terms of Human Activity Recognition (HAR), a task in which behavior means a sequence of pre-defined action classes such as running, walking, or sitting~\cite{Cristani2013}. Despite such a definition of behavior, HAR is a problem far from being solved~\cite{HAR2016} and still faces major challenges. For instance, different people can perform the same action in different ways~\cite{Bulling2014}, different actions can leave similar traces in a given sensor~\cite{Bulling2014,Kim2010}, there can be obtrusion, the number of people in a recording can change significantly (from an individual to crowds), the sensors can be inaccurate or the environmental conditions can change. Furthermore, only a few studies address HAR in a fully unsupervised way~\cite{Paoletti2020}, i.e., without knowing in advance what are the actions to be recognized. Additionally, a major limitation is that HAR approaches do not consider the social context in which humans perform actions and therefore they do not understand what such actions mean. For example, a HAR approach can understand that a person touches his/her own face by reasoning on the spatial relation between hands, face, and body parts~\cite{Beyan2020ICMI}. However, the same HAR approach does not necessarily try to understand whether self-touching results from thinking, being embarrassed, or mimicking another person (a behavior typically related to liking and rapport~\cite{Feese2012}). In this respect, current HAR approaches appear to lack \emph{Social Intelligence}, i.e., the ability to make sense of human behavior in socially meaningful terms.

The problem of interpreting observable behavior in social terms gave rise to \textit{Social Signal Processing} (SSP)~\cite{Pentland2007,Vinciarelli2009}, an interdisciplinary domain at the crossroad between technology and human sciences (psychology and sociology). In particular, SSP research integrates social psychology concepts into Artificial Intelligence (AI) and focuses on automatic detection, interpretation, and synthesis of \emph{social signals}, nonverbal behavioral cues (e.g., gestures, eye gaze, body posture, vocal characteristics, interpersonal distances, facial expressions) capable to convey socially-relevant information. The core idea underlying SSP and more generally, \emph{Social AI} (the collective name encompassing all domains aimed at endowing machines with social intelligence) is that observable human behavior is not just a sequence of actions but the physical, machine-detectable trace of social and psychological phenomena that cannot be sensed and observed directly. For example, from a Social AI point of view, a smile is not just the activation of a few facial muscles that move the lip corners up but the evidence that, with a certain probability, an individual is experiencing the emotion of happiness or is showing a friendly attitude. \\
\\
\noindent
\textbf{\textit{Scope.}}
The aim of this paper is to review the SSP-oriented approaches for HBU. We focus on \emph{nonverbal behavior analysis in co-located social interactions} and survey the research efforts tackling \emph{automated detection} of a \emph{social and psychological phenomenon}. Reviewed works are clustered into three: \textit{1)} detection of social traits, e.g., leadership, dominance, personality traits (Sec.~\ref{personalityLeaDom}), \textit{2)} social role recognition and social relations detection (Sec.~\ref{rolesSocialRelation}) and \textit{3)} interaction dynamics analysis to detect group cohesion, empathy, rapport and so forth (Sec.~\ref{groupdynamics}). Each study is discussed to provide answers to the following main research questions:
    \textit{1)} What are the nonverbal cues allowing to perform effective automated detection of a phenomenon?
    \textit{2)} What are the computational methodologies performing effective automated detection of a phenomenon through nonverbal signal representations? 
    
It is important to mention that even for the detection of the same social and psychological phenomena, the studies can differ from each other in terms of the sensor(s) they used, the interaction setting (e.g., dyads, small groups), and scenario (e.g., meetings, free-standing conversations) that their methodology was tested on. On the other hand, one should notice that such differences could determine the way automated interaction analysis is computationally performed through nonverbal signals. Therefore, our review also acknowledges the type of sensors used and specifies the interaction environment. We concentrate on the (technical) novelties and limitations of the reviewed works considering the time of the publication. For each phenomenon, we share our key observations and tend to highlight what has not yet been performed (e.g., the adaptation of deep models, and usage of multimodal cues). The further paper inclusion/exclusion criteria can be summarized as follows. Remote/online interactions, which have been getting more attention lately, are not in our scope. The interactants are always \emph{humans}, therefore, we do not concentrate on human-robot, -virtual agents, or similar interactions. All studies discussed rely on nonverbal signals and perform automated detection of any aforementioned social and psychological phenomenon. If a study also utilizes linguistic features together with nonverbal signals, that study was also considered as in the scope of this survey. Emotion recognition and synthesis (which are the topics of Affective Computing rather than SSP, see \cite{chanel2015connecting} for more information) are other important subjects, that can be realized through the analysis of nonverbal behaviors. However, this paper does not target such applications, unless their (or any other affective states') identification was addressed together with a social and psychological phenomenon. Besides, there exist several health-related applications (e.g., detection of depression \cite{alsarrani2022thin}, autism spectrum disorder \cite{georgescu2020reduced}), performed by automated nonverbal behavior analysis during the co-located human-human interactions. Such works are also out of our scope. Limited attempts to analyze interactions in images (i.e., spatial modeling only) were not considered because of the importance of modeling spatiotemporal data in HBU \cite{Hung2020}. Moreover, single-person settings, such as vlogs \cite{Teijeiro-Mosquera2015} or presentations (unless there is at least one other person who is co-located and being interacted with) are excluded. We consider the peer-reviewed research efforts published in the conference proceedings or journals (technical reports, and pre-prints are not covered) since 2010. The reader can refer to Appendix \ref{appendix1}, describing and illustrating the selection process of the papers reviewed in this work.\\
\\
\noindent
\textbf{\textit{Related Surveys in SSP Domain.}} This section summarizes prior related review papers in order to motivate the need for our survey paper and to demonstrate our differences and contributions with respect to them. One of the earliest surveys of nonverbal behavior (also called non-linguistic behavior) analysis was 
~\cite{Vinciarelli2009}. That work introduced, besides the concept of SSP, a taxonomy of nonverbal behavioral cues (grouped into physical appearance, gestures and posture, face and eye behavior, vocal behavior, and use of space and environment) and an initial indication of the social phenomena (i.e., personality, status, dominance, persuasion, regulation, and rapport) that it was possible to detect through automatic analysis of nonverbal cues. Furthermore, the survey showed how several technologies developed until then could be used for SSP purposes (e.g., speech analysis, computer vision, and biometry). In the same year, 
the studies about the automatic analysis of \emph{small group conversations} in terms of nonverbal communication were reviewed in \cite{GaticaPerez2009}. That work illustrated computational models of interaction management, internal states, personality traits, and social relationships. 
In parallel, 
the survey papers on nonverbal behavior analysis for \emph{video surveillance} \cite{Cristani2013} and
\emph{automatic personality perception, recognition, and synthesis} based on, e.g., written texts, nonverbal behaviors, mobile data, wearable devices, and online games \cite{Vinciarelli2014} was published. At the time the aforementioned surveys were released, most approaches were unimodal and based such as on the use of cameras or microphones to capture social interactions. In the meantime, the application of multimodal approaches has become more common, presumably, because the cost of wearable sensors keeps decreasing and devices such as smartphones and smartwatches (equipped with sensors such as cameras, microphones, accelerometers, and gyroscopes) have seamlessly integrated everyday life.
The use of multiple sensors, sensor networks, and multimodal methodologies brought novel perspectives and new findings. In particular, 
\cite{Palaghias2016} provided a review of \emph{mobile SSP applications} covering mobile sensing, social interaction detection, behavioral cues extraction, social signal inference, and social behavior understanding. Further significant improvements were raised which resulted from the recent advances in AI and especially from the development of deep learning methods.
In this respect, 
\cite{Stergiou2019} reviewed human-human interaction studies covering some of the SSP research in which \emph{Convolutional Neural Networks} (CNNs) recognize social behavior in videos. Furthermore, 
\cite{Mehta2020} focused on \emph{deep learning-based personality detection}. Other survey papers were; 
\cite{Junior2019} summarizing the progress in \emph{personality trait analysis in visual computing} and 
\cite{Rasipuram2020} focusing on automatic analysis of social interactions for \emph{soft skills prediction}. \\
\\
\noindent
\textbf{\textit{Our Contributions.}}
This paper has the following threefold contributions. \textit{1)} The types of social and psychological phenomena this review covers are much more than the context included by the related survey papers (notice that the research efforts on some phenomena, e.g., vocal entrainment, group performance, satisfaction, and quality of interactions have never been surveyed before). We also take into account all sensor technologies used and incorporate all possible interaction settings (e.g., dyads, small groups) and scenarios (e.g., meetings, free-standing conversations) in addition to discussing each paper in terms of the nonverbal cues and the computational methodologies. These make our survey majorly different from Junior et al. \cite{Junior2019} considering only cameras and visual domain, Cristani et al. \cite{Cristani2013} focusing only on video surveillance, Gatica-Perez \cite{GaticaPerez2009} concentrating on small group conversations, \cite{Junior2019} limiting its scope to the recognition of personality traits, and Mehta et al. \cite{Mehta2020} focusing only on deep learning methods. Moreover, given its submission date, our survey is able to include more recent literature. Please notice that several related surveys \cite{Vinciarelli2009,GaticaPerez2009,Cristani2013,Vinciarelli2014} were published before 2015. \textit{2)} We provide a review of the related datasets in terms of their scenario, group size, the total number of participants, annotations, sensors, and public availability information. Such a comprehensive summary is not available in the literature, and we argue that it can foster future efforts in data collection by helping to spot what has not been analyzed yet. Additionally, it can improve awareness regarding the datasets so that future studies can use them in their experimental analysis. \textit{3)} We propose several future research directions regarding dataset collection, privacy-preserving social interaction analysis, and the usage of Artificial Intelligence. Such proposals have not been introduced in any SSP-oriented paper before.
\\
\\
\noindent
\textbf{\textit{Organization.}} The rest of the paper is structured as follows. We first introduce the preliminaries in Sec.~\ref{preliminaries}, which are necessary to be able to make the review of existing literature. 
Our survey is given in Sec.~\ref{IntAnalysis}. In Sec.~\ref{datasets}, we examine the datasets having the annotations regarding the social and psychological phenomena inspected in Sec.~\ref{IntAnalysis}. Sec.~\ref{future} draws some possible future research directions motivated by challenges and the observations out of our review. Finally, Sec.~\ref{conclusion} summarizes the key observations, limitations, and future research directions.

\section{Preliminaries}
\label{preliminaries}
We first define the concepts of dyads and groups given that human-human social interaction in which interactants share the same environment is in the form of one of them (Sec.~\ref{socialDyadGroup}). 
We refer to multiple interaction environments, including \textit{structured settings} (such as meetings), \textit{semi-structured settings} (such as mingling gatherings), or \textit{in-the-wild settings} (for instance crowds). 
In Sec.~\ref{nonverbalCues}, we describe the nonverbal cues used by the reviewed studies. 
Sec.~\ref{sensors} reports the sensors utilized in co-located human-human social interaction analysis, particularly for the automated detection of social and psychological phenomena. Finally, Sec.~\ref{CompMet} exhibits the computational methods utilized by the reviewed papers.

\subsection{Dyads, Groups and Interaction Environments}
\label{socialDyadGroup}

The expression \emph{dyad} refers to two persons involved in social interaction while the expression \emph{group} refers to cases in which the number of interactants is at least three.
Dyad and group members feel like a part of a super-individual union and whenever collaborating towards results possibly subject to evaluation, share the responsibility of success~\cite{Hung2020}. Interaction takes place through both verbal and nonverbal communication affected by the number of interacting people. In particular, small groups (3-6 people) tend to be more dynamic than larger ones \cite{GaticaPerez2009}. For this reason, it is important to highlight that some studies consider dyads as groups of two persons, but dyads are different from groups in many aspects affecting the way automatic interaction analysis is performed \cite{GaticaPerez2009,Hung2020}. For instance, \textit{1)} people tend to form and dissolve dyads more often than groups, \textit{2)} dyad and group interactants experience different emotions, \textit{3)} the group members can interact with an individual person or a subgroup while dyad members can interact only with the other dyad member and furthermore, \textit{4)} some social phenomena (such as group cohesion) cannot be studied in dyads~\cite{Hung2020}. 

\begin{figure*}[ht]
	\centering
	\includegraphics[width=0.85\linewidth]{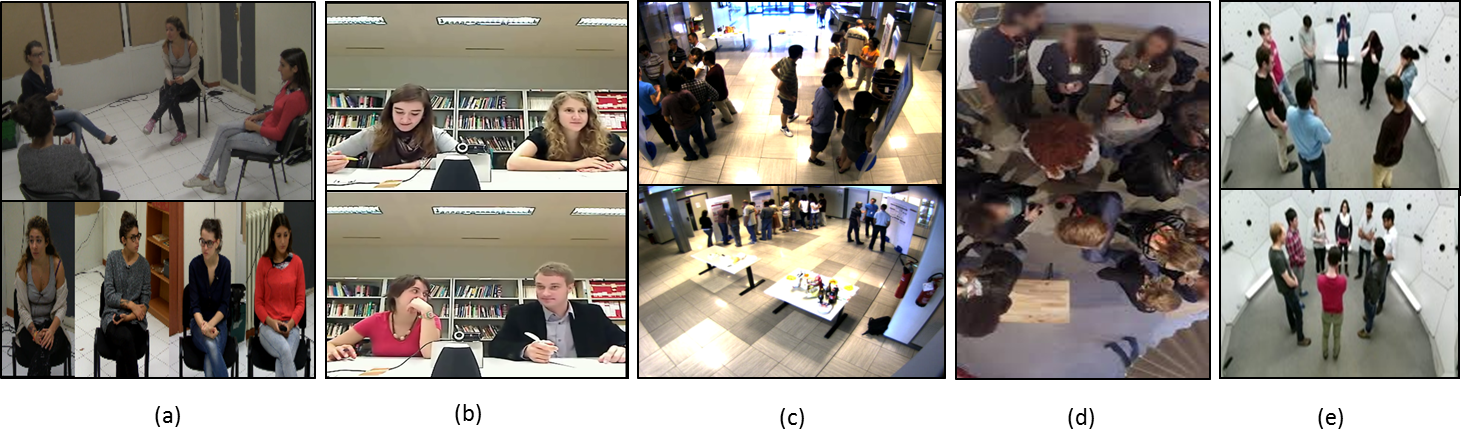}
	\vspace{-0.5cm}
	\caption{Scenes from some datasets: a) PAVIS Leadership \cite{Beyan2016} and b) ELEA \cite{SanchezCortes2012} investigate small group interactions when a group task is given, they address so-called meeting scenarios, c) SALSA \cite{AlamedaPineda2015} and d) MatchNMingle \cite{MatchNMingle} is composed of mingling events where the dyads/groups are dynamically formed without a predefined task to accomplish, and e) Panoptic \cite{Joo2019} includes standing participants who play a game, have pre-assigned roles implying that the group task is pre-defined.}
	\label{datasetFig}
	\vspace{-0.5cm}
\end{figure*}

The reviewed papers mostly focused on interactions in small groups, in particular the \emph{meetings} including 3-4 participants that sit around a table to perform a predefined task. Such a setting was used to infer a large number of social phenomena from nonverbal communication. Instead, fewer studies addressed the interactions in \emph{mingling events}~\cite{MatchNMingle} (also referred to as \emph{free-standing conversations}~\cite{AlamedaPineda2015}) involving a larger number of participants that dynamically (and even frequently) form dyads/groups and move from one dyad/group to the other without restrictions. People involved in such scenarios interact spontaneously and therefore the social interactions show more articulated and in-the-wild dynamics. It is important to notice that free-standing conversations might require the detection of groups before addressing the detection of social and psychological phenomena. This is different from the meetings that reviewed papers were set up in the way that dyad/group members are constant and hence performing automated group detection is not needed. There has been extensive literature on automatic group detection. In most cases, the focus was given to the detection of spatial arrangements and the most important cues were interpersonal distance and relative body orientation~\cite{Groh2010}. Furthermore, many works addressed the detection of \emph{F-formations}~\cite{Cristani2011b,Marquardt2012,Setti2013,Gan2013b,Ricci2015,Yanhao2015,Katevas2016,Kamper2017,Gedik2018,Rosatelli2019}, which are spatial arrangements that people spontaneously form during free-standing conversations. In the F-formations context, a person communicating with another tends to maintain close proximity and have a shared focus of attention. Most methodologies detecting F-formations used standard RGB videos captured by distant cameras \cite{Cristani2011b,Setti2013,Ricci2015,Yanhao2015,Inaba2016,Kamper2017,Varadarajan2018} while Gan et al. \cite{Gan2013b} utilized Kinect depth sensors as well. Alternatively, Katevas et al.
\cite{Katevas2016,Girolami2020} detected stationary interactions inside the crowds by relying on smartphones' Bluetooth and motion activity sensors. In 
\cite{Gedik2018}, 
pairwise F-formations using the data collected from a single body-worn tri-axial accelerometer were detected. On the other hand, there exist a few attempts to detect the social and psychological phenomena in crowds \cite{Kjargaard2014,Solera2015,Suzuki2013,Favaretto2019} such as by processing the indoor/outdoor surveillance systems' data. Fig.~\ref{datasetFig} presents some scenes taken from publicly available datasets demonstrating different interaction environments.

\subsection{Nonverbal Cues}
\label{nonverbalCues}

\begin{figure*}[t!]
	\centering
	\includegraphics[width=0.8\linewidth]{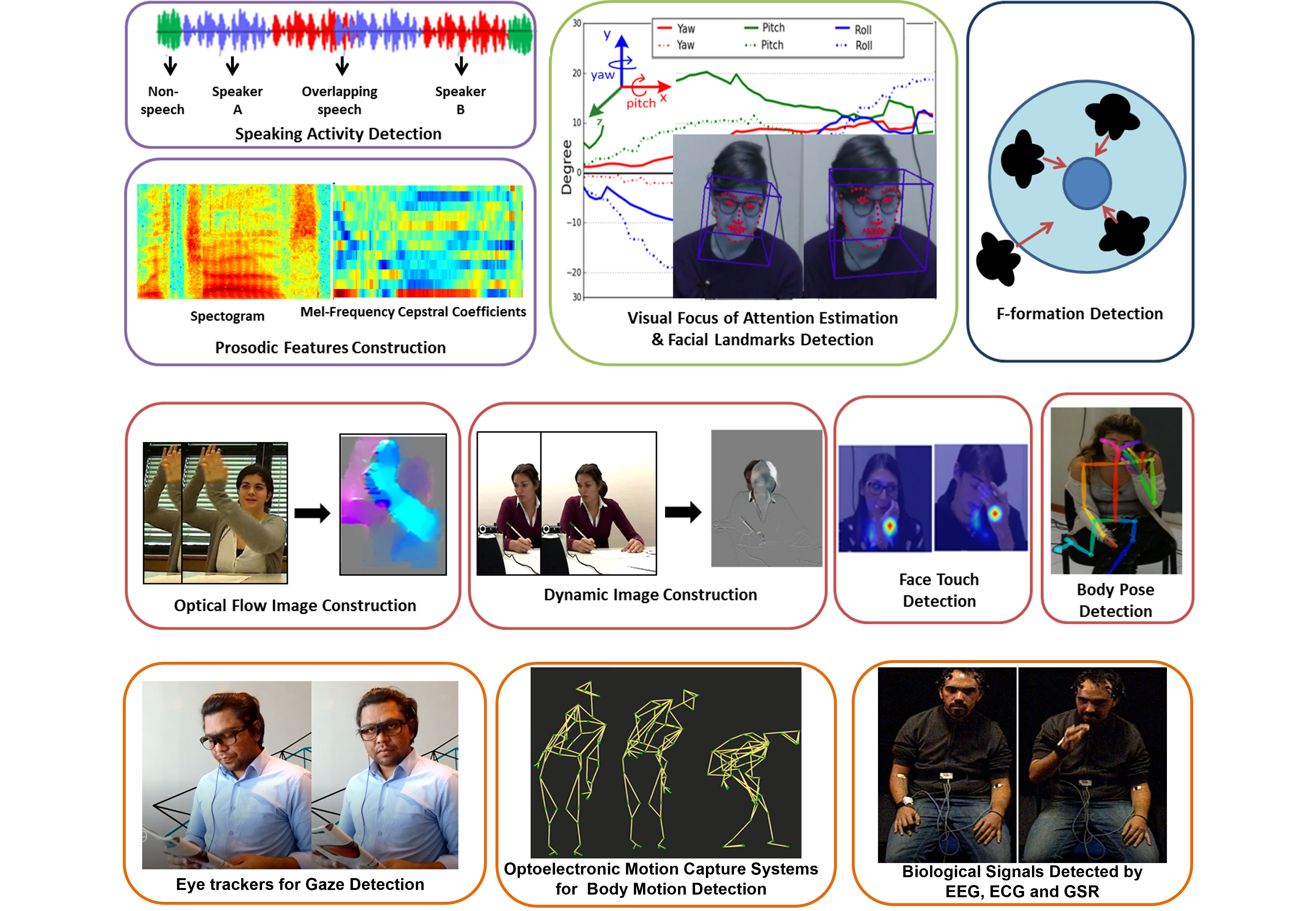}
	\vspace{-0.3cm}
	\caption{Some methodologies and sensors to extract nonverbal cues are used for the inference of various social and psychological phenomena during co-located human-human social interactions. The speaking activity detection and prosodic feature construction (e.g., spectrogram and Mel-frequency cepstral coefficients in image format) are for vocal behavior extraction. 
	The visual focus of attention estimation and facial landmarks detection (adapted from \cite{Beyan2016}) are for eye gaze-based nonverbal cues extraction and facial expression analysis, respectively. 
	F-formation detection can be used to detect social groups based on proximity. 
	Optical flow image construction (adapted from \cite{Beyan2018}), dynamic image construction (adapted from \cite{Beyan2019}), face-touch detection (adapted from \cite{Beyan2020ICMI}), and body pose detection (adapted from \cite{Beyan2017}) are the methods that were used to represent body motion and extract the associated nonverbal cues. 
	Eye trackers are to obtain the eye gaze of the wearer \cite{Thakur2021}, optoelectronic motion capture (MoCAP) systems are to detect the body motion of the wearer \cite{Niewiadomski2018} and electroencephalogram (EEG), electrocardiogram (ECG) and skin conductance sensor were used to capture biological signals of the wearer \cite{Correa2018}.
	}
	\vspace{-0.5cm}
	\label{NBCuesFigure}
\end{figure*}

The nonverbal cues used by the reviewed studies include \emph{vocal behavior} (Sec.~\ref{VocalBehavior}), \emph{body activity} (Sec.~\ref{BodyActivity}), \emph{eye gaze and visual focus of attention} (Sec.~\ref{EyeGaze}), \emph{facial expressions} (Sec.~\ref{Face}), \emph{proxemics} (Sec.~\ref{Proximity}) and \emph{physical appearance} (Sec.~\ref{PhysicalAppearance}).
Fig.~\ref{NBCuesFigure} illustrates some methodologies and/or sensors to automatically extract a number of nonverbal cues.

\subsubsection{Vocal Behavior}
\label{VocalBehavior}
Such behavior corresponds to everything in speech except words, including the use of vocalizations (e.g., fillers, laughter, and sobbing), pauses, and turn-taking. Speaking activity and prosody are the two aspects of vocal behavior that are most commonly used in the literature.\\

\noindent \textit{Speaking Activity.} 
The most common form of speaking activity detection is referred to as \emph{speaker diarization}, which is the automatic segmentation of speech recordings into \emph{turns} (i.e., time intervals) in which only one of the persons involved in an interaction is speaking. In other words, the speaker diarization detects \emph{who speaks and when}. If the identity of the speaker is not necessary, speaking activity detection can be performed by simply performing a speech/non-speech segmentation, i.e., by automatically identifying the intervals of time in which there is at least one person speaking. Both tasks above are typically addressed through audio processing methodologies, but there have been approaches that used alternative cues too (e.g., the movement of the speakers~\cite{SVVAD2020}). In most cases, the output of speaker diarization or speech/non-speech segmentation is represented in terms of statistical properties of turns (e.g., total length, mean, maximum, mean, interquartile range, histogram, fraction), speaking time length, overlapping speech time, turn-taking order, the number of successful/unsuccessful interruptions, and speaker floor grab, the fraction of time non-overlapping speech accounts for and the back-channel time during turns of different speakers. Such properties can be extracted for each individual involved in an interaction and/or at the group level. Examples of the latter case (typically called group conversational features) include group speaking time length and its statistics (e.g., total, mean) or the total number of conversational events (e.g., successful/unsuccessful interruptions, back-channel utterances). \\

\noindent \textit{Prosody.} Such cues account for the way people talk and in most cases correspond to how loud people speak (is captured through the energy of the speech signal) and their intonation (is captured through the fundamental frequency of the speech signal). In addition, prosody-related cues include speaking rhythm (captured by utterance timing). The features corresponding to these cues typically include statistics of signal energy, fundamental frequency (e.g., variation, maximum, mean), spectral features (formants, bandwidths, spectrum intensity), speaking rate (e.g., number of syllables per second), local variability of the speech signal (e.g., jitter and shimmer) and Mel-Frequency Cepstral Coefficients (MFCCs). 
Due to the large number of speech features aimed at capturing vocal behavior, there have been attempts to identify standard feature sets through the application of publicly available packages (e.g., OpenSMILE~\cite{Eyben2013}) or meta-analysis of the literature (e.g., the Geneva Minimalistic Acoustic Parameter Set (GeMAPS)~\cite{Zhong_interspeech_2019}).

\subsubsection{Body Activity}\label{BodyActivity}
The expression body activity refers to postures, movements of the upper body, arms, hands, and gestures. In most cases, the input data consists of videos and the first step is the detection of humans and/or their body parts, 
followed by the actual detection of the movements. The literature proposes publicly available packages (e.g., OpenPose \cite{Cao2019}) that can perform human body part detection, and their output is often used to represent movements with suitable terms for further use in the analysis of social interactions. There are multiple approaches to represent movements such as spatiotemporal processing of body pose~\cite{Beyan2017}, weighted motion energy images (wMEI)~\cite{SanchezCortes2012}, dynamic images~\cite{Beyan2019} and optical flow images~\cite{Beyan2018,Jinna2019}. Once the mentioned representations are available, it is possible to extract features such as statistics of movement's energy (e.g., mean and standard deviation), hand activity histograms, number of the body pose changes and variance of body orientation. An alternative approach is to apply data-driven techniques such as CNNs \cite{Beyan2019}.
Finally, 
object recognition and HAR were used to obtain the number of face-touching events, arm-folding, arm diagonal behavior, hand gesticulations \cite{Feese2012}, head/hand/arm/body fidgeting 
head nodding 
\cite{Oertel2015}, the interaction between people and a predefined set of objects and object manipulation~\cite{Ramanathan2013,Liu_2019_CVPR}.
\subsubsection{Eye Gaze \& Visual Focus of Attention}
\label{EyeGaze}
The Visual Focus of Attention (VFOA) can be defined as the point in space a person is looking at and it can be detected by estimating either gaze direction or head pose (some works, for example, \cite{Ramanathan2015}, use head and body pose as an indicator of attention). Compared to the head pose, the eye gaze direction tends to provide more accurate VFOA information. However, it is more difficult to estimate, and it requires one to constrain, at least to a certain extent, the movement of people. In particular, it requires the use of invasive sensors such as eye trackers that capture high-resolution images of the eyes but need to be worn and might not allow a person to freely move the head. For this reason, there are many studies (e.g., \cite{Beyan2016,Kindiroglu2017,Zhang2019,Subramanian2013}) that detected the VFOA by using the head pose. Such an approach was further encouraged by the availability of open-source packages aimed at head orientation measurement (e.g., OpenFace~\cite{openface}). The head pose corresponds to 3-angles (tilt or pitch, pan or yaw, and roll) that allow one to identify a vector normal to the face of a person. The direction of such a vector is used as an estimate of the gaze direction and leads to the identification of the VFOA. Once the VFOA of all people participating in an interaction is known, it is possible to calculate several features that were shown to account for different social and psychological phenomena. As examples of such features, the distribution of the time when an individual is in the VFOA of others, the amount of time two people are in the VFOA of each other at the same moment, the amount of time the VFOA corresponds to a given object and the number of times the VFOA of an individual corresponds to another one can be given.

\subsubsection{Facial Expressions}\label{Face}
%
Facial expressions were specifically included in the analysis of dyads (e.g., interviews) or meetings. Their usage was relatively less than speaking activity, prosody, eye gaze/VFOA, and body activity such that the number of studies covering them is seven (which are \cite{Muller2019,Chongyang2019,Yanbang2020,Srivastava2012,Naim2015,Muralidhar2018,ZhangRadke2020}) whereas the number of works analyzing the speaking activity, prosody, eye gaze/VFOA, and body activity is 53, 47, 32 and 49, respectively. Some phenomena whose automated detection was performed based on facial expressions are; rapport \cite{Muller2018}, leadership \cite{Muller2019,Yanbang2020}, dominance \cite{Chongyang2019}, personality traits \cite{Srivastava2012} and social roles \cite{ZhangRadke2020}). Facial expression analysis includes detecting facial landmarks and facial action units (FAUs). A popular tool to extract such information is OpenFace \cite{openface} providing 68 facial landmarks and 17 distinct FAUs for a face typically captured by a close-up camera. 
\subsubsection{Proxemics}\label{Proximity}
This is the domain studying the social meaning of interpersonal distances especially when the movement of people is not constrained and therefore social, cultural, and psychological phenomena are the only factors underlying the physical distance between two individuals~\cite{riosmartinez2015}. Free-standing conversations (conv.) are the settings in which proxemics play the most important role and individual distances were shown to account for the detection of multiple social phenomena including social roles~\cite{AlamedaPineda2015}, leadership~\cite{Suzuki2013, Kjargaard2014}, engagement~\cite{Veenstra2011} and personality traits~\cite{Subramanian2013,Subramanian2013,CabreraQuiros2016b,Favaretto2019,Dotti2020}.
The sensors used to detect the proximity were various from cameras to smartphones and smart badges. In the literature, a social group corresponds to individuals exhibiting similar motion characteristics such as having similar trajectories, being close to each other, or having similar motion orientation. A convenient way to define social groups is by detecting F-formations 
as described in the previous section.
\subsubsection{Physical Appearance}\label{PhysicalAppearance}
This includes height, body shape, skin/hair color, clothes, and make-up of a person. It was addressed in terms of physical attractiveness, sympathy, and appreciation in~\cite{Nguyen2014}. Besides, the appearance of a person and contextual objects are explored to distinguish different social relations in~\cite{Liu_2019_CVPR}. By the time deep learning methods (especially CNNs) have been used in the context of this paper, physical appearance is (implicitly) being considered more, by performing feature learning from RGB images. One corresponding implementation can be seen in~\cite{ZhangRadke2020}.
\subsection{Sensors}
\label{sensors}
Sensors and their arrangement determine the way automatic interaction analysis is performed. In particular, the cues that can be detected depend on the sensors being used for data acquisition, and correspondingly, the robustness of cue detection changes significantly with the modalities and sensors used. For what concerns vocal behavior, the most frequent sensors are microphones of different types: close-talk (CTM), headset (HSM), tabletop (TTM), lapel (LAM), microphone arrays (ARM), omnidirectional (ODM), four-channel cardioid (FCM) and other distantly placed microphones such as single-channel far-field (FFM). In the case of body activity, eye gaze, VFOA, facial expressions, proximity, and the physical appearance of the participants, the main sensors are cameras. Depending on the setting and scenario, the literature shows the use of cameras: close-up (CUC), Red Green Blue Depth (RGBD), web (WC), 360 degrees (C360), frame-synchronized (FSC), digital movie (DMC), top-view (TVC), wearable (WCM), narrative (NC), pan and tilt zoom (PTZ), far-field (FFC) or a combination of them. CUC has been used mostly in meeting environments, for instance, in~\cite{Muller2019,Beyan2019TMM} for soft skills analysis and in~\cite{Othman2019} for social group performance analysis whereas TVC has been utilized in proximity analysis such as in~\cite{Veenstra2011}.

Although the most common sensors are microphones and cameras, other sensing technologies have also attracted researchers. For example, smartphones were the preferred sensors in \cite{Flutura2016} as they represent the core communication device in daily life. In fact, smartphones can be used to efficiently monitor co-located human-human interactions in dyads and groups because they are equipped with a wide array of sensors such as microphones, cameras, accelerometers (ACC), gyroscopes (GYRO), magnetometers (MAG) and proximity (Prox) detectors. In a similar vein, sociometer badges (SB) are small portable devices that include microphones, infrared beam detectors, Bluetooth detectors, and ACC sensors. They were used for the analysis of free-standing conversations in \cite{AlamedaPineda2015,Kalimeri2011,Dotti2020,Dotti2020b,Jayagopi2012b,Yanxia2018,Zhang2018}. Eye trackers (ET) and smart glasses (SG) have been very precise devices for capturing visual attention, in particular for dyadic \cite{Terven2016} or standing interactions \cite{Suzuki2013}. Recently, they were also used during rather more complex co-located interactions in which the wearer moves a lot and interacts with multiple objects/humans in indoor and outdoor environments \cite{Thakur2021}. External ACC and Prox were used for the analysis of mingling scenarios \cite{CabreraQuiros2016b} while ACC was used for the analysis of small group round-table meetings \cite{Nihei2014} as well. Wearable inertial measurement units (IMU) \cite{Feese2011,Lahnakoski2020,Thakur2021} and motion capture systems (MoCAP) \cite{Suzuki2013,Nihei2014,Niewiadomski2018} were typically preferred to realize highly robust body motion analysis.
Head-mounted wearables were used to capture the speech, body movement, and gaze \cite{hadley2019speech}. As an example, \cite{li2020think} showed that eye activity determined by head-mounted wearable could be used to infer a person's communication load and conversational state that is to be applied for turn-taking prediction. Other wearables such as respiration sensors \cite{Ishii2014} were shown to be useful for predicting the next speaker in co-located multi-party meetings. Biological sensors combined with sensors capturing nonverbal behaviors also supply an accurate automated understanding of various social phenomena. For example, \cite{klados2020automatic} realized automatic personality recognition on a dataset \cite{Correa2018} captured by electroencephalogram (EEG) and annotated based on the observations of the nonverbal behaviors.

\subsection{Computational Methods}
\label{CompMet}
The computational methods considered in this survey refer to 
machine learning or deep learning supervised and unsupervised approaches. While supervised methods were frequently utilized, pioneer works on some topics such as modeling leadership \cite{Sanchez2010,SanchezCortes2012,SanchezCortes2012b} and dominance \cite{Aran2010b,Hung2011} applied unsupervised methods that are rule-based or ranking-based or based on the fusion of features. Another popular unsupervised method was Granger causality, a multivariate time series representation framework applied in \cite{Kalimeri2011,Kalimeri2012} for modeling dominance. Finally, Gaussian Mixture Models (GMM) and Latent Dirichlet Allocation (LDA), further unsupervised approaches, were applied for engagement analysis in \cite{Oertel2013} and leadership style detection in \cite{Jayagopi2010}. The supervised methods were used for performing classification or regression. The most popular classification method was Support Vector Machine (SVM), which was typically combined with a radial basis function (RBF) kernel or a linear kernel. Earlier works used Naive Bayes (NB) \cite{Feese2011,Jayagopi2012b,Matic2014,Lahnakoski2020,Hayley2010,Nanninga2017,Hagad2011}, Locally Weighted Naive Bayes (LW-NB) \cite{Subramanian2013} and k-Nearest Neighbors (k-NN) \cite{Veenstra2011} as well. Bayesian Networks (BN) \cite{Kalimeri2010,Vinciarelli2011}, Boosting \cite{Valente2012,Kubasova2019}, Boosted Trees (BT) \cite{Murray2018}, Decision Tress (DT) \cite{Zhang_2020_WACV}, Random Forest (RF) \cite{Chongyang2019,Subramanian2013,Kindiroglu2017,Nguyen2014,Okada2019TMCCA,Nihei2014,Murray2018,Kubasova2019,Othman2019,Zhang2018}, Collective Classifier (CC) \cite{SanchezCortes2012} and Multiple Kernel Learning (MKL; and its variations such as Localized MKL (LMLK)) \cite{Beyan2017,Beyan2019TMM,Beyan2019,Beyan2018} combined with other classifiers such as SVM, were other popular methods. Some of these classifiers were merged within Multi-task Learning (MTL) \cite{Kindiroglu2017}.
In case of sequential data exist, the most commonly applied methodologies include Influence Model (IM)~\cite{Staiano2011b}, Hidden Markov Models (HMM) - including variations such as Linear-chain HMM (LC-HMM)~\cite{Staiano2011b} - and Conditional Random Fields (CRF) - including variants such as linear-chain CRF (LC-CRF)~\cite{Staiano2011b,Sapru2013,Sapru2015,Fotedar2016,Ramanathan2013}. A similarly wide spectrum of approaches can be observed in the case of regression methods that include linear regression (linR), Support Vector Regression (SVR) \cite{Go2019,Lai2018}, Random Forest Regression (RFR) \cite{Lai2018}, Ridge Regression (RR) \cite{Srivastava2012,Aran2013b,Aran2013,Okada2015,Fang2016,Kindiroglu2017,Nguyen2014,Okada2019TMCCA}, Multiple Regression (MR) \cite{Suzuki2013,Bhattacharya2018} and Logistic Regression (LR) \cite{CabreraQuiros2016b,Jinna2018,Kapcak2019,Nihei2014,Zhang2018}.

The latest years have witnessed the success of Deep Neural Networks (DNN) that were increasingly more often used, from shallow models such as Multilayer Perceptron (MLP) \cite{Chongyang2019,Hagad2011} to deep architectures such as Convolutional Neural Networks (CNN) \cite{Dotti2020b,Beyan2018,Jinna2018}, Fully Connected Neural Networks (FCNN) \cite{ZhangRadke2020} and Graph Convolutional Networks (GCNN) \cite{Lin2020}. Deep Boltzmann Machines (DBM) were applied as an unsupervised feature learning method to jointly model body pose features in \cite{Beyan2017}. For sequential data processing, Long Short-Term Memory Networks (LSTM) \cite{Beyan2019,Lin2018,ZhangRadke2020,Jinna2019,Aimar2019,Zhong_interspeech_2019,Sharma2019}, a type of Recurrent Neural Network (RNN), were applied for various problems and often in combination with CNNs. In \cite{Beyan2019TMM}, sequential data processing was performed with Conditional Restricted Boltzmann Machines (CRBM) \cite{Beyan2019TMM} and a combination of RNNs with Restricted Boltzmann Machines (RNN-RBM) \cite{Beyan2019TMM}. On the other hand, 
\cite{Palmero_2021_WACV} presented a Transformer-based context-aware model to analyze face-to-face dyadic interactions.
The next section discusses in depth the computational methods and particularly their effectiveness together with the nonverbal cues utilized.

When considering the computational methods, it becomes essential to emphasize their evaluation, including the specific evaluation metrics employed.
In this regard, there are no distinct evaluation metrics beyond the conventional ones used in standard classification (e.g., accuracy, f-score, recall, precision) and regression (e.g., mean square error) evaluations, which are applicable to various machine and deep learning methods. Consequently, there is a lack of standardization across studies regarding evaluation metrics. It is worth noting that accuracy is the most commonly used metric for classification tasks, while mean square error is frequently utilized for regression analysis.

\section{Review: Methodologies to Detect Social and Psychological Phenomena}
\label{IntAnalysis}
Recalling that, this paper reviews the studies performing automated detection of social and psychological phenomena, which are realized through automatic nonverbal behavior analysis of co-located human-human social interactions, we divided the existing literature into three broad categories. These are namely social traits (Sec.~\ref{personalityLeaDom}), social roles or relations (Sec.~\ref{rolesSocialRelation}), and interaction dynamics (Sec.~\ref{groupdynamics}). Below, we discuss each study in text and further summarize them in Tables \ref{personalityLeaDom_table}, \ref{roleTable}, and \ref{groupdynamics_table}. The tables are recommended to be read together with the corresponding text. Our discussions start with a brief definition of social and psychological phenomena. We then focus on the effectiveness of the nonverbal cues and the computational methods applied along with the corresponding sensors and the interaction environment.


{\small\tabcolsep=3pt  
\begin{center}
\footnotesize
\begin{longtable}{|p{0.6cm}|p{2.5cm}|p{3.8cm}|p{2.8cm}|p{4cm}|}

\caption{Summary of the research efforts performing automatic detection of the social traits: personality traits, leadership, dominance, competence, liking, and hirability. The references discussed for each category are listed chronologically. See text for abbreviations.} \label{personalityLeaDom_table} \\  

\hline \multicolumn{1}{|l|}{\textbf{Ref.}} & \multicolumn{1}{l|}{\textbf{Sensor}} &
\multicolumn{1}{l|}{\textbf{Nonverbal Cue}} &
\multicolumn{1}{l|}{\textbf{Scenario (Group Size)}} &
\multicolumn{1}{l|}{\textbf{Computational Method}} \\ \hline
\endfirsthead

\multicolumn{5}{c}%
{{\bfseries \tablename\ \thetable{} -- continued from previous page}} \\
\hline \multicolumn{1}{|l|}{\textbf{Ref.}} & \multicolumn{1}{l|}{\textbf{ Sensor}} &
\multicolumn{1}{l|}{\textbf{ Nonverbal Cue }} &
\multicolumn{1}{l|}{\textbf{ Scenario (Group Size)}} &
\multicolumn{1}{l|}{\textbf{ Computational Method }} \\ \hline 

\endhead

\hline \multicolumn{5}{|r|}{{Continued on next page}} \\ \hline
\endfoot

\hline \hline
\endlastfoot

\multicolumn{5}{l}{\textbf{(Emergent) Leadership}} \\ 
\rowcolor{LightCyan}
\cite{Sanchez2010} & ARM & Speaking activity & Meetings (3-4) & Rule-based  \\ 
\rowcolor{LightCyan}
\cite{SanchezCortes2012} & ARM, WC & Speaking/head/body activity, prosody & Meetings (3-4) & Rule-based, rank-based fusion, SVM, CC \\ 
\cite{SanchezCortes2012b}  & ARM, WC & Speaking activity, gaze & Meetings (3-4) & Rank-based fusion \\ 
\rowcolor{LightCyan}
\cite{Suzuki2013} & IMU, CTM, ET, MoCAP & Eye gaze, movement, proximity & Standing (6) & Stepwise MR  \\ 
\cite{Kjargaard2014} & MP & Proximity & Indoor, crowd & PageRank-based graph link analysis  \\
\rowcolor{LightCyan}
\cite{Solera2015} & TVC & Speed, position, acceleration & Crowd & Structural SVM \\ 
\cite{Beyan2016} & CUC & VFOA & Meetings (4) & Rank-based fusion, SVM  \\ 
\rowcolor{LightCyan}
\cite{Beyan2016b} & CUC & VFOA, head/body activity & Meetings (4) & SVM, MKL \\ 
\cite{Beyan2017} & LAM, CUC & Posture, VFOA, speaking/head/body activity & Meetings (4) & DBM w/ \{SVM, LMKL\} \\ 
\rowcolor{LightCyan}
\cite{Bhattacharya2018} & Kinect, LAM & VFOA, speaking activity & Meetings (3-5) & Linear MR \\ 
\cite{Niewiadomski2018} & MoCAP & 3D body pose, eye gaze & Dance (dyad) & Rule-based  \\ 
\rowcolor{LightCyan}
\cite{Muller2019} & LAM, CUC, FSC, CTM & VFOA, body pose, FAU, speaking activity & Meetings (3-4) & SVM w/rbf and Domain Adaptation  \\ 
\cite{Beyan2019TMM} & LAM, CUC & Speaking activity, VFOA & Meetings (4) & CRBM, RNN-RBM w/\{SVM, LMKL\}  \\ \hline  
\multicolumn{5}{l}{\textbf{Leadership style}} \\
\rowcolor{LightCyan}
\cite{Jayagopi2010} & HSM & Speaking activity & Meetings (4) & LDA \\ 
\cite{Feese2011} & IMU & Body posture & Meetings (3) & NB  \\ 
\rowcolor{LightCyan}
\cite{Beyan2017b} & LAM, CUC & Speaking/head/body activity, prosody, VFOA & Meetings (4) & SVM, MKL \\ \hline 
\multicolumn{5}{l}{\textbf{Dominance}} \\ 
\rowcolor{LightCyan}
\cite{Aran2010b} & HSM, LAM, ARM, CUC & Speaking/head/body activity & Meetings (4) &  Rank-based fusion \\ 
\cite{Jie2010} & HSM, LAM, ARM & Speaking activity, energy & Meetings (4) & Rule-based  \\
\rowcolor{LightCyan}
\cite{Escalera2010} & HSM, WC & Speaking/body/face/mouth activity & Dyadic interactions & Adaboost \\
\cite{Hung2011} & OTM & Speaking activity & Meetings (4) & Rule-based\\ 
\rowcolor{LightCyan}
\cite{Kalimeri2011} & SB & Body movement energy & Meetings (4) & Granger Causality \\ 
\cite{Kalimeri2012} & CTM, CUC & Speaking energy, body activity & Meetings (4) & Granger Causality \\ 
\rowcolor{LightCyan}
\cite{Chongyang2019} & ARM, WC & Speaking activity, VFOA, FAU, emotion scores, MFCCs & Meetings (3-4, 5-8) & MLP, RF\\ 
\cite{Yanbang2020} & ARM, WC & MFCCs, VFOA, FAU, speaking activity, emotion intensity & Meetings (3-4) & Temporal network-diffusion CNN \\ \hline 
\multicolumn{5}{l}{\textbf{Personality traits}} \\ 
\rowcolor{LightCyan}
\cite{Lepri2010} & CTM, ODM, CUC, WC & Conversational activity, emphasis, influence, mimicry, fidgeting & Meetings (4) & SVM w/ rbf \\ 
\cite{Lepri2010W} & CTM, ODM, CUC, WC & Speaking activity, VFOA & Meetings (4) & SVM w/ rbf \\ 
\rowcolor{LightCyan}
\cite{Kalimeri2010} & CTM, ODM, CUC, WC & Conversational activity, emphasis, influence, mimicry, fidgeting & Meetings (4) & BN \\  
\cite{Staiano2011} & CTM, ODM, CUC, WC & Conversational activity, emphasis, VFOA & Meetings (4) & SVM w/\{linear, rbf\} \\ 
\rowcolor{LightCyan}
\cite{Staiano2011b} & CTM, ODM, CUC, WC & Conversational activity, emphasis, VFOA & Meetings (4) & IC-HMM, LC-CRF, IM   \\ 
\cite{Valente2012} & HSM & Speaking activity, prosody & Meetings (4) & Boosting \\ 
\rowcolor{LightCyan}
\cite{Srivastava2012} & DMC & Speaking activity, prosody, facial expressions & Movies & RR, SVM w/\{linear, sparse, rbf\} \\
\cite{Lepri2012} & CTM, ODM, WC & Speaking activity, VFOA & Meetings (4) & SVM w/rbf  \\ 
\rowcolor{LightCyan}
\cite{Aran2013b} & WC & Visual activity & Meetings (3,4) & RR, linear SVM\\ 
\cite{Aran2013} & ARM, WC & Speaking/head/body activity, prosody, VFOA & Meetings (3-4) & RR, linear SVM  \\ 
\rowcolor{LightCyan}
\cite{Subramanian2013} & FSC & VFOA, proximity & Indoor free-standing conv. (6) & Linear SVM, RF, LW-NB  \\ 
\cite{Okada2015} & ARM, WC & Speaking / body activity, prosody, VFOA & Meetings (3-4) & RR, linear SVM \\ 
\rowcolor{LightCyan}
\cite{CabreraQuiros2016b} & ACC, Prox & Speaking activity, body movement energy, proximity & Mingle events (30-32) & LR \\ 
\cite{Beyan2019} & WC & Dynamic images \cite{Bilen2016} & Meetings (3-4) & CNN+LSTM w/\{softmax, SVM, LMKL\} \\ 
\rowcolor{LightCyan}
\cite{Favaretto2019} & TVC & Proximity, velocity, trajectory angular variation & Indoor/outdoor surveillance (crowd) & Rule-based \\
\cite{Dotti2020} & FSC, SB & Body pose, proximity & Indoor free-standing conv. (18) & Metric learning \\ 
\rowcolor{LightCyan}
\cite{Dotti2020b} & FSC, SB & Body pose, proximity, social group motion & Indoor free-standing conv. (18) & CNN \\  
\rowcolor{LightCyan}
\cite{Palmero_2021_WACV} & LAM, CUC, ODM & Face, raw video, raw audio
& Dyadic interactions & 2D-CNN, R(2+1)D network, Transformer \\ \hline 
\multicolumn{5}{l}{\textbf{Personality traits together with leadership/dominance/competence/liking detection}} \\ 
\rowcolor{LightCyan}
\cite{Fang2016} & ARM, WC & Speaking / head / body activity & Meetings (3-4) & RR, SVM w/ rbf   \\ 
\cite{Kindiroglu2017} & ARM, WC & Speaking / head / body activity, VFOA, prosody & Meetings (3-4) & SVM, RR, RF, MTL w/ regression \\ 
\rowcolor{LightCyan}
\cite{Lin2018} & ARM & Speaking activity & Meetings (3-4) & Bidirectional LSTM  \\ 
\cite{Beyan2018} & LAM, ARM, WC & Optical flow images & Meetings (3-4) & CNN w/ \{softmax, SVM, LMKL\} \\ 
\rowcolor{LightCyan}
\cite{Zhang_2020_WACV} & CUC & VFOA, body activity and hand-head / face relative position &  Meetings (3-4) & Multiple DT \\ \hline 
\multicolumn{5}{l}{\textbf{Hirability}} \\ 
\rowcolor{LightCyan}
\cite{Nguyen2014} & ARM, CUC & Speaking/body/head activity, prosody, head nods, smile, eye gaze, physical appearance & Real job interviews (dyad) & Ordinary least-squares, RR, RF \\ 
\cite{Naim2015} & CUC & Prosody, facial expressions & Mock interviews (dyad) & SVR, Lasso, LR, Gaussian Process \\ \hline
\multicolumn{5}{l}{\textbf{Hirability together with other soft skills}} \\ 
\rowcolor{LightCyan}
\cite{Muralidhar2018} & Kinect, ARM & Speaking activity, prosody, body motion, head nods, facial expressions,  & Job interviews and conv. in reception desk (dyad) & SVR, RFR \\
\cite{Okada2019TMCCA} & ARM, WC & Speaking/head/body activity, prosody, VFOA & Meetings (3-4), job interviews (dyad) & RR, linear SVM, RF\\ \hline 

\end{longtable}
\normalsize
\end{center}
}
\vspace{-1cm}
\subsection{Social Traits}
\label{personalityLeaDom}
The social traits addressed by the computing studies include competence, likeability, hirability, personality, dominance, and leadership. The task is the automatic prediction of such traits based on the self-assessments or the external observers’ judgments. This section defines each trait and then discusses the approaches, findings, novelties, and limitations of each relevant work. Furthermore, a detailed summary of reviewed papers can be seen in Table~\ref{personalityLeaDom_table}.\\

\noindent \textbf{\textit{Leadership and its style.}} A leader is a person who has authority and power over a group of people and can exert his/her dominance, influence, and control over them \cite{Beyan2016}. Emergent leaders, on the other hand, are the ones who naturally show these characteristics in a group. 
Reviewed works investigated the leadership and its style for their automated detection.

Emergent leadership has been studied comprehensively to classify a person as the most leader, the least leader, or other \cite{Sanchez2010,SanchezCortes2012,SanchezCortes2012b,Beyan2016,Beyan2016b,Beyan2017,Muller2019,Beyan2019TMM}.
As the interaction environment, small group meetings, composed of 3-4 participants sitting around a table, were preferred the most \cite{Sanchez2010,SanchezCortes2012,SanchezCortes2012b,Beyan2016,Beyan2016b,Beyan2017,Bhattacharya2018,Muller2019,Beyan2019TMM}. There exist only a few study performing detection of the designated leaders (also called role-based leaders) such that \cite{Kjargaard2014,Solera2015,Suzuki2013} focusing on crowds and \cite{Niewiadomski2018} targeting the interactions among dyads. Overall, the most frequently used modalities were audio and video, which were captured by ARM/LAM and CUC, respectively. Differently, Kinect was preferred in \cite{Bhattacharya2018}, IMU and ET were used in \cite{Suzuki2013}, and MoCAP was utilized in \cite{Suzuki2013,Niewiadomski2018}. The nonverbal signal and the computational method employed the most were; speaking activity and SVM, respectively. Sanchez et al. \cite{Sanchez2010,SanchezCortes2012,SanchezCortes2012b} presented the first publicly available dataset for emergent leadership. Additionally, they deeply examined the effectiveness of several feature sets based on speaking activity, head and body activity, prosody, and gaze together with several learning models and proposed a novel method called Collective Classifier. Besides, the first study investigating the leadership in standing groups composed of six persons was \cite{Suzuki2013}, showing the effectiveness of eye gaze, movement, and proximity modeled with stepwise MR. The works \cite{Kjargaard2014,Solera2015} focused on crowds and presented novel approaches based on PageRank-based graph link analysis and structural SVM, respectively. The analysis in \cite{Solera2015} was limited to the overall performance of the model without further discovering the individual contribution of each feature. Several studies \cite{SanchezCortes2012,SanchezCortes2012b,Muller2019,Beyan2019TMM} showed that using multimodal features improves inference performance. For example, it was found that visual information augments acoustic information \cite{SanchezCortes2012}. Beyan et al. \cite{Beyan2016b} demonstrated that VFOA and speaking activity-based features can perform better than head and body activity-based cues. Furthermore, effective joint modeling of multimodal cues was studied in \cite{Beyan2016,Beyan2017,Beyan2019TMM}, indicating the state-of-the-art (SOTA) performance of MKL methods. Even though earlier art applied rule-based or rank-based models \cite{Sanchez2010,SanchezCortes2012,SanchezCortes2012b,Kjargaard2014}, deep architectures have been also applied in \cite{Beyan2017,Beyan2019TMM}. Notice that, there is not yet an end-to-end pipeline. Pose-based visual activity modeling was prominent in \cite{Beyan2017,Suzuki2013,Niewiadomski2018,Muller2019}. For instance, a novel methodology encoding the spatiotemporal body postures within an unsupervised learning approach was presented in \cite{Beyan2017}. That method became the SOTA body activity representation for emergent leadership. As the only attempt, Muller et al. \cite{Muller2019} performed an analysis across different leadership datasets, testing the generalizability of an SVM trained on one dataset to others, showing that cross-dataset leader detection is possible by using pose and VFOA features.

Regarding leadership style detection, Jayagopi et al. \cite{Jayagopi2010} investigated autocratic leaders, participative leaders (egalitarian), or free-rein leaders (those who allow group members to make decisions) by analyzing the participants' speaking activity in meetings. The focus was given to designated leaders in~\cite{Jayagopi2010}. Instead, Beyan et al. \cite{Beyan2017b} aimed to detect emergent leaders' style, which was classified either as autocratic (a person who is directive, tends to exert strong control and decision making) or democratic (a person who tends to involve group members into decision-making process). Also, a wider set of cues including speaking activity, prosody, head/body activity, and VFOA were modeled within unimodal and multimodal learning schemes in \cite{Beyan2017b}. The demonstrated results were more favorable towards the speaking activity and VFOA \cite{Beyan2017b}. Differently, Feese et al. \cite{Feese2011} discriminated between individually considerate leaders (defined as leaders who pay attention to their followers and listen to them effectively) and autocratic leaders using data from IMU sensors. That has been the unique work \cite{Feese2011} showing the effectiveness of body posture mirroring that was modeled with NB. However, it was limited due to not presenting comparisons with other nonverbal signals. Overall, one can observe that there has been no attempt to perform leadership style prediction using deep models.
\\
\\
\noindent \textbf{\textit{Dominance.}} It is seen as a trait or it is the hierarchical position of a person in a group \cite{Kalimeri2012}. As a trait (the case of our interest), dominance characterizes people that influence others and the automatic detection of the dominant person in a dyad/group was the aim of the reviewed studies.
Following social psychology, dominance leaves traces in terms of speaking time, turn management, interruptions, pitch, body activity, facial expression, VFOA, and eye gaze. For this reason, computing studies have focused mostly on such nonverbal cues \cite{Aran2010b,Jie2010,Escalera2010,Hung2011,Kalimeri2011,Kalimeri2012,Chongyang2019,Yanbang2020}. The most widely investigated interaction setting was meetings composed of 3-4 people \cite{Aran2010b,Jie2010,Hung2011,Kalimeri2011,Kalimeri2012,Yanbang2020}. Instead, Escalera et al. \cite{Escalera2010} examined dyads, and Chongyang et al. \cite{Chongyang2019} studied larger groups of up to eight people. The usage of the microphones (HSM, KAM, ARM), and cameras (CUC, WC) hold the majority while the wearable SB was preferred in \cite{Kalimeri2011}.

The research efforts in the dominance context mainly focused on improving automatic detection performance. To do so, the studies proposed different nonverbal cue combinations with peculiar computational models. Overall, the speaking activity cues are the most effective ones while the body activity cues follow them. The deep models in \cite{Chongyang2019,Yanbang2020} boosted the SOTA performance significantly. Aran et al. \cite{Aran2010b} applied multimodal fusion of speaking, head, and body activity cues at the feature extraction level and at the classifier level via score and rank level fusion. The results showed that the visual information is complementary to audio and multimodal fusion is needed to achieve better dominance estimation performance. Jie et al. \cite{Jie2010} focused on speaking length and speaking energy to recognize the most and the least dominant persons in multi-party conversations. The simple rule-based method was found effective compared to SVMs. However, applying feature extraction in a semi-automatic way is a limitation of \cite{Jie2010}. Escalera et al. \cite{Escalera2010} presented a set of movement-based features extracted from body, face, and mouth activities in order to define a higher set of interaction indicators that were modeled with AdaBoost and demonstrated that the speaking length is the preferred feature to detect dominant people. Motivated by this finding, Hung et al. \cite{Hung2011} examined how the performance of a speaker diarization method affects the dominance estimation and found that reducing the signal-to-noise ratio of the input source and reducing the computational complexity of the speaker diarization algorithm both leads to worse performance. 
The novelty of \cite{Kalimeri2011,Kalimeri2012} lies in using Granger Causality, which took into account the causal effects the dominant subjects' speaking and body activities have on the behavior of the other participants. Chongyang et al. \cite{Chongyang2019} not only estimated the most dominant person in a group but also inferred the more dominant person in a pair of people by proposing the Dominance Rank algorithm. That study \cite{Chongyang2019} proved the effectiveness of the facial action units, emotion scores (the intensity of eight emotions and two facial traits (smile, and open eyes)), MFCCs, speaking and gazing probabilities modeled with an MLP. Yanbang et al. \cite{Yanbang2020} used the features' of \cite{Chongyang2019} and presented a better-performing temporal network. That model's effectiveness is thanks to containing a diffusion block capable of extracting patterns from highly dynamic interactions and being able to learn subtle and local patterns by integrating a set of poolings.
\\
\\
\noindent \textbf{\textit{Personality Traits.}} Personality is a latent construct that accounts for individuals' behavior, characteristics, emotions, and motivations~\cite{Vinciarelli2014,Mehta2020}. The reviewed computing studies focused on personality traits for automatic recognition. Different personality models were proposed in social psychology. However, in technological research \cite{Staiano2011,Staiano2011b,Valente2012,Aran2013b,Aran2013,Okada2015,Beyan2019,Favaretto2019,Dotti2020,Dotti2020b} the most popular model has been the Big-Five traits, which correspond to \emph{openness to experience} (the person who is curious, original and have wide imagination and interest), \emph{conscientiousness} (the person who is efficient, organized, reliable, responsible and thoughtful), \emph{extraversion} (the person who is active, energetic, talkative and assertive), \emph{agreeableness} (the person who is trustworthy, straightforward, kind, forgiving and generous) and \emph{neuroticism} (the person who is nervous, tense, sensitive, unstable and worrying). Big-Five traits have been often referred to as OCEAN traits because of their initials. Overall, extraversion was the most often investigated trait as it can be strongly displayed during group interactions and consequently resulted in more reliable annotations than other personality traits \cite{Aran2013}. On the other hand, \emph{Locus of Control} (LoC) is defined as the outcomes of a person's actions are primarily the results of her own actions or based on the events/influences outside of his/her control \cite{Rotter1966GeneralizedEF} and it was examined in \cite{Kalimeri2010,Lepri2010}. HEXACO traits include honesty in the Big-Five cluster and they were studied in \cite{CabreraQuiros2016b}.

The automatic recognition of these traits was repeatedly performed with speaking activity, VFOA, prosody, and body activity-based cues. Three nonverbal behavior groups were constructed based on the categories described in Sec.~\ref{nonverbalCues} in~\cite{Lepri2010,Lepri2010W,Kalimeri2010,Staiano2011,Staiano2011b}. These were: \emph{i) conversational activity} which is an indicator of interest and engagement consists of the audio energy in a time frame, length of the voiced segments, length of the speaking segments, the fraction of the speaking time, and voicing rate; \emph{ii) emphasis} referring to the speaker‘s motivation such that the consistency can be interpreted as focus and its variability can be interpreted as openness, including the prosodic features formant frequency, confidence in formant frequency, spectral entropy, values of the larger autocorrelation peaks, number of the larger auto-correlation peaks, location of the larger auto-correlation peaks, and time derivative of energy and \emph{iii) influence} that is related to dominance and represented by the ratio of overlapping speech segments to the total segments. The majority of the works preferred investigating the meetings involving 3-4 people \cite{Lepri2010,Lepri2010W,Kalimeri2010,Staiano2011,Staiano2011b,Valente2012,Lepri2012,Aran2013b,Aran2013,Okada2015,Beyan2019}. Consequently, the most frequently used sensors were CTM, ARM, ODM, CUC, and WC. Some others tested their method, which uses speaking activity, prosody, and facial expressions, on a dataset composed of movies \cite{Srivastava2012}, in free-standing conversations using interpersonal distances \cite{Subramanian2013,CabreraQuiros2016b,Dotti2020,Dotti2020b} and in indoor/outdoor surveillance videos where people engage in social gatherings \cite{Favaretto2019}. Palmero et al. \cite{Palmero_2021_WACV} focused on dyadic interactions and used facial features, raw audio, and several metadata (interlocutors’ characteristics, order of task, task difficulty, interlocutors’ relationship).

The most important results and findings are summarized as follows. For extraversion and LoC identification, using activity and emphasis features together improved the performance \cite{Lepri2010}. Speaking activity and gaze features were effective in predicting extraversion, and the classification performance depended on the length of the observation time \cite{Lepri2010W}. Extraversion was the best-recognized personality quality which was detected by using vocal behavioral cues (such as pitch) and VFOA in \cite{Staiano2011}. Following that, openness was the second best-recognized quality, and the corresponding result was obtained when the speech energy derivative was used. The usage of IM resulted in better performance than HMMs and became an alternative approach to perform Big-Five traits detection \cite{Staiano2011b}. Non-linguistic features (prosody, speech activity, overlaps, and interruptions) could outperform linguistic features (words n-gram and dialog acts) for the prediction of high/low extraversion, consciousness, and neuroticism traits \cite{Valente2012}. Okada et al. \cite{Okada2015} proposed the usage of higher-level features representing inter-modal and inter-person relationships, whose patterns are identified as co-occurring events based on a clustering algorithm. This novel representation of features demonstrated improved results compared to earlier works. The way features were fused (early fusion) in \cite{Okada2015} was altered to late fusion in a follow-up paper of authors \cite{Okada2019TMCCA}, resulting in even better performance. Aran et al. \cite{Aran2013b} used vlog datasets as the source and the face-to-face small group meetings as the target to train a ridge regression model, which was later tested on the target dataset. As the target data used was labeled, that approach contradicts unsupervised domain adaptation, still, it has been contributing to the literature by showing the generalizability of the body activity cues across datasets. Later on, the effectiveness of using thin-sliced impressions was compared with the whole meeting impressions of external observers in \cite{Aran2013}. The results implied that for the extraversion trait, predictions based on the whole meeting are feasible with a slight decrease compared to the predictions based on the thin-sliced data. Whereas for the openness trait, using the whole meeting provides better performance. Moreover, body activity features could achieve higher accuracy for extraversion detection in comparison to audio features, and the performance of energy features was found to be higher than other prosodic features \cite{Aran2013}. VFOA features were good at predicting extraversion and neuroticism while the head pose errors had more impact on extraversion detection in cocktail party scenarios in \cite{Subramanian2013}. The data of body-worn triaxial ACC to extract the individual's speaking status was merged with movement energy-related cues to estimate the personality in crowded mingling scenarios in \cite{CabreraQuiros2016b}. That study \cite{CabreraQuiros2016b} relied on transductive parameter transfer across speech-related body movements of some participants to label the speech turns of others. 
Beyan et al. \cite{Beyan2018} proposed a novel pipeline composed of optical flow image computation (a representation of body activity), CNN-based feature learning, feature encoding performed through covariance matrices in Riemann space and the classification with SVM and LMKL. Such an approach \cite{Beyan2018} showed significantly better results compared to the prior visual activity-based and audio-visual features for high/low extraversion classification. Instead of optical flow images, dynamic image representation \cite{Bilen2016} was the input of a CNN+LSTM network aiming to detect the spatiotemporal saliency to determine key-dynamic images in \cite{Beyan2019}. As a novelty, only the feature embeddings belonging to the automatically detected test key-dynamic images were used for personality prediction and such an implementation improved the performance compared to using all images in a test video. Another study  \cite{Dotti2020} proposed an architecture composed of two CNNs, jointly modeled with a triplet loss. The novel inputs of that model were images embedding the skeleton data over time and the interpersonal distance representing the global dynamics of a scene over time. That proposed method \cite{Dotti2020} surpassed several prior arts including LSTM modeling body motion and proxemics features over time. The same model was extended in \cite{Dotti2020b} by adding the social group descriptor in the image format and a third CNN-based encoder, resulting in better scores than \cite{Dotti2020}.
The facial features extracted from the person-in-interest and the scene features extracted from other interlocutors were used together with the raw audio signal in \cite{Palmero_2021_WACV}. That study was the first work using Transformers for personality trait recognition. However, using several CNNs as the backbones made the whole pipeline computationally heavy.

Among all social phenomena, personality traits have gathered the highest attention from computing studies. Deep learning approaches are the SOTA for this research line. However, it is hard to make a conclusion on the effectiveness of a computational model over another as the studies were rarely tested on the same datasets. Due to this fact, the comprehensive highlights are exhibited only in terms of nonverbal cues as follows. Speaking activity and prosody-based features were found effective for the automatic recognition of several personality traits. As an alternative, body activity features have been very promising even in semi-structured scenarios such as poster sessions (a type of free-standing conversation). There exist several recent efforts representing the body activity over time in terms of images to process them with CNNs \cite{Beyan2018,Beyan2019,Dotti2020,Dotti2020b}. There is no doubt that multimodal (e.g., acoustic and vision) approaches dominate this research line due to their remarkably better performance than unimodal methods.
\\
\\
\noindent \textbf{\textit{Competence \& Liking.}}  
A person who is perceived as \emph{competent} has the experience, skills, knowledge, and intelligence while a person who is perceived as \emph{liking} is kind, friendly, sympathetic, considerate, and/or well-disposed \cite{SanchezCortes2012}. 
Such social traits were examined together with leadership and/or personality traits by computing studies.
The entire efforts shown for the automated detection of such traits undertook the interactions during meetings composed of 3-4 persons \cite{Fang2016,Kindiroglu2017,Lin2018}. Thereupon, the sensors used were bounded with microphones (ARM, LAM) and cameras (WC, CUC). Fang et al. \cite{Fang2016} studied perceived competence, liking, dominance, leadership, and the rank of dominance together with Big-Five personality traits. The effectiveness of the speaking activity, energy and pitch, body motion, and audio interruptions were investigated with regression and SVM. As the novelty, that study \cite{Fang2016} approached the problem in terms of:
intra-personal features (i.e. related to only one participant), dyadic features (i.e. related to a pair of participants), and one-vs-all features (i.e. related to one participant versus the other members of the group). The intra-personal and one-vs-all features were found more effective for all traits and social impressions considered. Kindiroglu et al. \cite{Kindiroglu2017} also studied the same traits and social impressions. Similar to \cite{Aran2013b} a vlog dataset was transferred to a small group meeting dataset, but different from \cite{Aran2013b}, \cite{Kindiroglu2017} applied multi-task learning. Such an implementation resulted in improved performance with a limited size of target data. Lin et al. \cite{Lin2018} presented a novel attention mechanism (called interlocutor-modulated) injected into a bi-directional LSTM. That method modeled the vocal behaviors of both the target speaker and his contextual interlocutors and showed better performance than \cite{Fang2016,Kindiroglu2017}. The key conclusions are: \textit{i)} the dominant effectiveness of the speaking activity-based cues is prominent and \textit{ii)} the proposed method of \cite{Lin2018} is the current SOTA.
\\
\\
\noindent \textbf{\textit{Hirability.}}
It is defined in terms of communicative competence, persuasion skills, work conscientiousness, and stress resistance \cite{Okada2019TMCCA}. As expected, this trait was investigated during interviews (i.e., dyadic conversations) but also in small group meetings. Both settings were equipped with microphones (i.e., ARM) and cameras (i.e., Kinect, WC, CUC). The automatic detection of hirability is a social trait in which the facial expressions (as well as higher level cues inferred from facial expressions such as smile) were relatively more considered \cite{Naim2015,Muralidhar2018}. Speaking, body and head activities (particularly head nods), and VFOA were the other features examined. Studies used rather standard regression methods (e.g., SVR and LR), and the deep learning models have never been tested. The important findings can be encapsulated into three folds. \textit{i)} Multimodal nonverbal cues perform better than unimodal audio cues (speaking activity, prosody) or unimodal video-based features (VFOA, head, and body activities) \cite{Nguyen2014,Okada2019TMCCA}.
\textit{ii)} Out of the above-mentioned cues, the most predictive ones are the applicant's audio cues and the interviewer's visual cues. This shows that the interviewer-produced responses could condition the behavior of the job applicant, i.e., by displaying visual back-channels \cite{Nguyen2014}. \textit{iii)} The higher speaking turn duration and head nods are positively correlated to higher ratings for hirability, competence, sociability, persuasiveness, clearness, and positiveness while verbal content is less important \cite{Muralidhar2018}.


{\small\tabcolsep=3pt  
\begin{center}
\footnotesize
\begin{longtable}{|p{0.6cm}|p{2cm}|p{6cm}|p{2.8cm}|p{2.8cm}|}

\caption{The summary of the computing works addressing automated social role recognition and social relations detection. The references discussed for each category are listed chronologically. See text for abbreviations.} \label{roleTable} \\

\hline \multicolumn{1}{|l|}{\textbf{Ref.}} & \multicolumn{1}{l|}{\textbf{Sensor}} &
\multicolumn{1}{l|}{\textbf{Nonverbal Cue}} &
\multicolumn{1}{l|}{\textbf{Scenario (Group Size)}} &
\multicolumn{1}{l|}{\textbf{Computational Method}} \\ \hline
\endfirsthead

\multicolumn{5}{c}%
{{\bfseries \tablename\ \thetable{} -- continued from previous page}} \\
\hline \multicolumn{1}{|l|}{\textbf{Ref.}} & \multicolumn{1}{l|}{\textbf{ Sensor}} &
\multicolumn{1}{l|}{\textbf{ Nonverbal Cue }} &
\multicolumn{1}{l|}{\textbf{ Scenario (Group Size)}} &
\multicolumn{1}{l|}{\textbf{ Computational Method }} \\ \hline 

\endhead

\hline \multicolumn{5}{|r|}{{Continued on next page}} \\ \hline
\endfoot

\hline \hline
\endlastfoot

\multicolumn{5}{l}{\textbf{Social Role Recognition}} \\ 
\rowcolor{LightCyan}
\cite{Vinciarelli2011} & HSM, LAM & Speaking activity, prosody & Meetings (4) & Dynamic BN \\ 
\cite{Sapru2013} & HSM, LAM & Speaking activity, prosody & Meetings (4) & CRF \\ 
\rowcolor{LightCyan}
\cite{Dong2013} & CTM, ODM, WC& Speaking activity, hand/body fidgeting & Meetings (4) & IM \\ 
\cite{Sapru2015} & HSM & OpenSMILE acoustic features \cite{Eyben2013}, speaking activity & Meetings (4) & CRF \\ 
\rowcolor{LightCyan}
\cite{Oertel2015} & CTM, Kinect & Speaking activity, visual backchannels, VFOA & Meetings (4) & SVM w/ RBF \\ 
\cite{Fotedar2016} & HSM & OpenSMILE acoustic features \cite{Eyben2013}, speaking activity & Meetings (4) & HCRF \\ 
\rowcolor{LightCyan}
\cite{AlamedaPineda2015} & FSC, SB & Head / body pose & Indoor free-standing conv. (18) & Alternating Direction Optimization\\
\cite{flemotomos2018combined} & FFM & MFCCs &  Motivational interview (dyad) & Hierarchical Agglomerative
Clustering, SVM \\
\rowcolor{LightCyan}
\cite{ZhangRadke2020} & HSM, CUC & Facial expressions, head pose, eye gaze, body movement, appearance, prosody, and audio short-term metrics \cite{pyAudioAnalysis} & Meetings (4) & LSTM+FCNN\\ \hline 
\multicolumn{5}{l}{\textbf{Social Relations Detection}} \\ 
\rowcolor{LightCyan}
\cite{Ramanathan2013} & DMC & Body motion, human-object int., facial images & YouTube videos & CRF \\ 
\cite{Jinna2018} & DMC & Spectrograms, motion in RGB and optical flow & Movies & CNN w/LR \\ 
\rowcolor{LightCyan}
\cite{Jinna2019} & DMC & Motion in images, optical flow, face images & Movies & CNN w/LSTM \\ 
\cite{Aimar2019} & WCM (NC) & RGB images & Indoor/outdoor free-sitting conv. & CNN+LSTM \\
\rowcolor{LightCyan}
\cite{Liu_2019_CVPR} & DMC & Physical appearance, proximity, human-object int. & Movies & Pyramid GCN \\ \hline 

\end{longtable}
\normalsize
\end{center}
}

\vspace{-1cm}
\subsection{Social Roles and Relations}
\label{rolesSocialRelation}
Below, we present a deep review of computing studies on automated social roles and social relation detection, also summarized in Table~\ref{roleTable}. \\

\noindent \textbf{\textit{Social Roles.}} 
The roles considered in studies published before 2010 typically correspond to specific functions \emph{role-played} in a given social context (such as a project manager in a meeting). Therefore, they are scenario-specific. However, a model trained in a dataset having a fixed social context might not be generalizable and subsequently might not be successfully applied to other datasets or real-world environments. Considering this, Vinciarelli et al. \cite{Vinciarelli2011} defined the following \emph{socio-emotional roles} that are independent of any particular scenario or corpus and several works \cite{Sapru2013,Dong2013,Sapru2015,Fotedar2016,ZhangRadke2020} proposed methods to detect such roles automatically.
\emph{Protagonist} is a person who takes the floor, leads the conversation, and asserts his/her authority, \emph{supporter} is a person who is cooperative, shows attention and acceptance while providing support, \emph{neutral} is a person who does not express his/her ideas and accepts others' ideas, \emph{gatekeeper} is the moderator who encourages the communication between the group members and \emph{attacker} is a person who does not agree with others' ideas, does not respect the status of others and attacks other speakers. Alameda-Pineda et al. \cite{AlamedaPineda2015} defined \emph{social attractor} (a person who attracts the attention of the other members in the group) to be detected in a real-world workshop session where people have discussions in front of posters. In such a context, the poster presenters were labeled as social attractors. Oertel et al. \cite{Oertel2015} instead studied the automated detection of \emph{attentive listener} (a person who grabs the floor after the current speaker), \emph{side participant} (potential future speakers that do not grab the floor after the current speaker) and \emph{bystander} (a person who is not expected to speak in the near future but the rest of the group is aware of them).

The automated recognition of social roles was largely addressed in meetings composed of 4-persons \cite{Vinciarelli2011,Sapru2013,Dong2013,Sapru2015,Oertel2015,Fotedar2016,ZhangRadke2020}. Few studies focused on other settings such as indoor free-standing conversations \cite{AlamedaPineda2015} and dyads \cite{flemotomos2018combined}. Microphones (HSM, LAM, CTM, ODM) were frequently incorporated to capture the relevant data followed by the cameras (WC, Kinect) and the usage of SB in free-standing conversations. The integration of such sensors implies that vocal behavior (speaking activity and prosody) was relatively well-investigated in this respect, which indeed can be observed in \cite{Vinciarelli2011,Sapru2013,Dong2013,Sapru2015,Oertel2015,Fotedar2016,ZhangRadke2020,flemotomos2018combined}. An exception is Alameda-Pineda et al. \cite{AlamedaPineda2015} which relied only on head and body postures. Others utilized a multimodal cue set such that the facial expressions, head pose, eye gaze, body movement, appearance, and prosody were modeled in \cite{ZhangRadke2020} while in addition to speaking activity Oertel et al. \cite{Oertel2015} and Dong et al. \cite{Dong2013} involved visual backchannels and hand/body fidgeting, respectively. The core findings of the reviewed works are summarized as follows. The effectiveness of turn-taking patterns, turn duration, and prosodic features with Bayesian BN modeling was shown in  \cite{Vinciarelli2011}. Sapru et al. \cite{Sapru2013,Sapru2015} applied a novel and more effective method than \cite{Vinciarelli2011}. They combined turn-taking patterns, speech duration, prosody, and lexical information to integrate the statistical dependencies between roles across adjacent segments of meetings with CRFs.
For the first attempt, in \cite{Dong2013} the importance of the temporal dependencies among \textit{a)} the roles played by the same subject, \textit{b)} the time properties of the roles played by each individual, and \textit{c)} the mutual constraints among the roles of different group members were considered through an IM framework, showing notable results. Fotedar et al. \cite{Fotedar2016} relied on a generative model where the participants' role transition was used and the likelihood of the feature vector for a role was generated by CRF. Such an approach showed improvements in the detection of roles; gatekeeper and protagonist compared to earlier works. One limitation of that study is assuming the availability of all slices from a meeting for every participant and therefore not being resilient to missing information. Oertel et al. \cite{Oertel2015} showed that speaking activity, visual/verbal backchannels, and gaze patterns observed in dyadic interactions remain the same in multi-party interactions during the detection of the role of silent participant. Such a conclusion is very interesting, still, the results should have been confirmed on a larger dataset as well as better examine the effect of data segment size on the findings. Recent work presented the usage of co-occurrence features and successive occurrence
features in thin time windows to model the behavior of a
person, as well as the responses of that person by using multi-stream RNN \cite{ZhangRadke2020}. That pipeline has been original for role detection and resulted in several conclusions such that: \textit{a)} VFOA is the most effective cue out of others and facial action units: lip-tightener, lip-tightener, lip-stretcher, and lip-corner-depressor are more descriptive among all action units, \textit{b)} the head yaw has the largest importance among all head orientations, \textit{c)} rhythm of the speech plays an important role while the entropy of energy, speaking status, spectral entropy and spectral roll off are among the important cues \cite{ZhangRadke2020}. The optimization procedure; alternating-direction method of multipliers (ADMM) was first time applied for role recognition in \cite{AlamedaPineda2015}. Such a model can be favorable for the estimation of any phenomena by solving the matrix completion problem with ADMM, independent of the nonverbal cues considered. Flemotomos et al. \cite{flemotomos2018combined} demonstrated that in dyads, using speech signals yields superior results compared to the independent use of turn-level classifiers which do not take speaker-specific variabilities. However, the experimental analysis was performed only with hierarchical agglomerative clustering and SVM. \\
\\
\noindent \textbf{\textit{Social Relations.}} Automated detection of social relations was majorly addressed based on computer vision techniques. The used sensors are different types of cameras and some of them recorded the audio too. The datasets composed of movies  \cite{Jinna2018,Jinna2019,Liu_2019_CVPR} or YouTube videos \cite{Ramanathan2013} were used for the validation of the proposed methods. One exception is \cite{Aimar2019} investigating indoor/outdoor free-sitting conversations. Ramanathan et al. \cite{Ramanathan2013} recognized social relations during events such as a birthday party where the relations are birthday person, parents, friends, and guests or a wedding where the roles are bride, groom, priest, and bridesmaids by modeling person-specific features; body motion, face images and human-object interactions through a customized CRF. The novelty of that work \cite{Ramanathan2013} was taking into account the inter-role interactions in CRF, showing considerably better performance with respect to k-means and baseline CRF (i.e., not considering the inter-role interactions). Social relations were split into subjective relations (dominant, competitive, trusting, warm, friendly, etc.) and objective relations (working relation, kinship relation, etc.) in \cite{Jinna2018,Jinna2019}. A multi-stream CNN was proposed to model audio and high-level semantic information in videos. The feature embeddings were further fused with LR in \cite{Jinna2018} while \cite{Jinna2019} exploited LSTM with multiple attention units, eventually outperforming LR. Liu et al. \cite{Liu_2019_CVPR} recognized social relations: attachment (relations: parent-offspring roles), mating (relations: couple), hierarchical power (relations: leader-subordinate, service (e.g., customer-waiter, passenger-driver), reciprocity (relations: sibling, friend) and coalitions groups (relations: colleague, opponent) with a framework called Multi-scale Spatial-Temporal Reasoning (MSTR). MSTR adapts temporal segment network (TSN \cite{wang2016temporal}), a triple graph model representing the visual relations between persons and objects as well as proposing Pyramid Graph Convolutional Network performing temporal reasoning with multiscale receptive fields. That method \cite{Liu_2019_CVPR} demonstrated better results than several complex architectures: TSN with/without spatiotemporal features, GCN, and its variations. Differently, a method to recognize social relations from egocentric video streams (notice that this has been the unique work processing first-person videos for social relation detection) composed of several interactions such as father-child, mother-child, lovers, colleagues were presented in \cite{Aimar2019}. Features learned from full, facial, or full-body images with various CNNs were compared with an LSTM-based classifier. Overall, several efforts in this context targeted integrating the advances of deep learning methods and even brought in novel technical perspectives, promoting more accurate social relation detection.


{\small\tabcolsep=3pt  
\begin{center}
\footnotesize
\begin{longtable}{|p{0.6cm}|p{2.2cm}|p{5.5cm}|p{2.8cm}|p{3cm}|}

\caption{The summary of the works performing interaction dynamics analysis to detect group conversational context, engagement, involvement, interest level, group performance, group satisfaction, group cohesion, vocal entrainment, rapport, and empathy. The references discussed for each category are listed chronologically. See text for abbreviations.} \label{groupdynamics_table} \\

\hline \multicolumn{1}{|l|}{\textbf{Ref.}} & \multicolumn{1}{l|}{\textbf{Sensor}} &
\multicolumn{1}{l|}{\textbf{Nonverbal Cue}} &
\multicolumn{1}{l|}{\textbf{Scenario (Group Size)}} &
\multicolumn{1}{l|}{\textbf{Computational Method}} \\ \hline
\endfirsthead

\multicolumn{5}{c}%
{{\bfseries \tablename\ \thetable{} -- continued from previous page}} \\
\hline \multicolumn{1}{|l|}{\textbf{Ref.}} & \multicolumn{1}{l|}{\textbf{ Sensor}} &
\multicolumn{1}{l|}{\textbf{ Nonverbal Cue }} &
\multicolumn{1}{l|}{\textbf{ Scenario (Group Size)}} &
\multicolumn{1}{l|}{\textbf{ Computational Method }} \\ \hline 

\endhead

\hline \multicolumn{5}{|r|}{{Continued on next page}} \\ \hline
\endfoot

\hline \hline
\endlastfoot

\multicolumn{5}{l}{\textbf{Group Conversational Context Classification}} \\
\rowcolor{LightCyan}
\cite{Jayagopi2012b}& SB & Speaking activity & Meetings (4) & NB, linear SVM \\ 
\cite{Matic2014} & MP, ACC & Speaking activity, proximity, body orientation & Workplaces, in-the-wild & NB, SVM \\ 
\rowcolor{LightCyan}
\cite{Bano2017} & GoPro Hero4 & Face, speaking activity, motion & In-the-wild int. ($>$2) & SVM \\ 
\cite{Miller2019} & HSM & Speaking activity & Meetings (3-5) & Situated data mining, Parallel episodes \cite{Patnaik2008} \\ \hline
\multicolumn{5}{l}{\textbf{Engagement / Involvement / Interest Level Detection}} \\
\rowcolor{LightCyan}
\cite{Veenstra2011} & TVC & Position, proximity, motion & Speed-dates (dyad) & SVM, k-NN \\
\cite{xiao2012multimodal} & CUC & Prosody (vocal energy), body motion, body pose & Conv. on conflictual topics (dyad) & SVM \\
\rowcolor{LightCyan}
\cite{bednarik2012gaze} & CUC & Eye gaze (fixations, saccades) & Meeting (4) & SVM w/ rbf \\
\cite{Oertel2013a} & CUC & Eye gaze & Meetings (8) & GMM \\ 
\rowcolor{LightCyan}
\cite{wang2013automatic} & HSM, FFM & Prosody (MFCCs, spectral cues, energy, speaking duration, shimmer, jitter) & Interview about a product (dyad) & SVM w/ rbf \\
\cite{kawahara2013estimation} & Eye-tracker, MoCAP & Eye gaze, gaze duration & Free-standing conv. (3) & NB \\
\rowcolor{LightCyan}
\cite{huang2017speaker} & CUC, TTM & Prosody,
facial features (energy image, HoG, Gabor) 
& Natural conv. (dyad) & bi-directional LSTM \\
\cite{Kapcak2019} & ACC & Movement & Speed-dates (dyad) & LR \\ 
\rowcolor{LightCyan}
\multicolumn{5}{l}{\textbf{Group Performance Prediction}} \\
\rowcolor{LightCyan}
\cite{Nihei2014} & HSM, WC, ACC & Prosody, head movement, eye gaze & Meetings (4) & SVM, RF, LR \\ 
\cite{Avci2014} & ARM, WC & Speaking/head/body activity & Meetings (3-4) & IM  \\ 
\rowcolor{LightCyan}
\cite{Avci2016} & ARM, WC & Speaking activity, body motion, VFOA & Meetings (3-4) & IM \\ 
\cite{Murray2018} & ARM & OpenSMILE acoustic features \cite{Eyben2013} & Meetings (3-4) & Gradient BT, RF, DNN \\  
\rowcolor{LightCyan}
\cite{Yanxia2018} & SB & Proximity & In-the-wild (6) & LDA\\ 
\cite{Zhong_interspeech_2019}  & LAM & GeMAPS \cite{Eyben2016} & Meetings (3) & DNN, bi-directional LSTM \\
\rowcolor{LightCyan}
\cite{Go2019} & HSM, WC, ACC & Speaking activity, prosody, head movement & Meetings (4) & SVR  \\  
\cite{Kubasova2019} & ARM, OTM & OpenSMILE acoustic features \cite{Eyben2013} & Meetings (2-4) & RF, Boosting, Extra Trees \\  
\rowcolor{LightCyan}
\cite{Othman2019} & CUC, HSM & Head pose, FAU &  Meetings (4) & RF \\ 
\cite{Lin2020} & ARM, OTM, LAM & OpenSMILE acoustic features \cite{Eyben2013} &  Meetings (3-4) & GCNN \\ \hline 
\multicolumn{5}{l}{\textbf{Group Satisfaction Detection}} \\
\rowcolor{LightCyan}
\cite{Lai2018} & HSM & Speaking activity, OpenSMILE features \cite{Eyben2013} & Meetings (4) & Bayesian RR, RFR, SVR \\ 
\cite{johnson2021clustering} &  WC, handy cam & Speaking activity  
& Meetings (2-3) 
& k-means clustering \\ \hline 
\multicolumn{5}{l}{\textbf{Estimating Quality of
Social Interactions}} \\
\rowcolor{LightCyan}
\cite{Lahnakoski2020} & Kinect, IMU & Proximity, face/body orientation, VFOA & Meetings (dyad) & NB, SVM \\
\hline
\multicolumn{5}{l}{\textbf{Vocal Entrainment Detection}}
\\
\rowcolor{LightCyan}
\cite{lee2010quantification} & ARM & Prosody (pitch, energy) & Couples' int. (dyad) & Statistical sequence modeling \\ 
\cite{georgiou2011behavioral} & ARM & Prosody (pitch, energy) & Couples' int. (dyad) & SVM \\ 
\rowcolor{LightCyan}
\cite{lee2012using} & ARM & Prosody (pitch, intensity), speech rate, MFCCs & Couples' int. (dyad)  & Correlation analysis \\ \hline
\multicolumn{5}{l}{\textbf{Group Cohesion Estimation}} \\
\rowcolor{LightCyan}
\cite{Hayley2010} & HSM, CUC & Speaking activity, prosody, head / body motion & Meetings (4) & NB, SVM \\ 
\cite{Nanninga2017} & HSM & Speaking rate and frequency & Meetings (3-8) & NB \\  
\rowcolor{LightCyan}
\cite{Zhang2018} & SB & Speaking activity, energy and consistency of the movement & Meetings (6) & LR, linear SVM, RF \\ 
\cite{Sharma2019} & NA & OpenSMILE acoustic features \cite{Eyben2013}, appearance & Youtube videos & CNN, LSTM \\ \hline 
\multicolumn{5}{l}{\textbf{Rapport / Empathy Detection}} \\
\rowcolor{LightCyan}
\cite{Hagad2011} & CUC & Posture & Meetings (dyad) & NB, SVM, MLP \\ 
\cite{park2012already} & CUC & Head nod/shake/tilt, eye gaze, smile, self-touch & Meetings (dyad) & SVM  \\
\rowcolor{LightCyan}
\cite{song2012multimodal} & CUC, LAM  & Head nod / shake, forefinger raise / raise-like / wag, hand wag/scissor, shoulder shrug, prosody 
& Political debates (5) & Canonical Correlation w/ Multi-view Hidden CRF \\
\cite{xiao2013modeling} & FFM & Prosody (pitch, jitter, vocal energy, shimmer) & Motivational Interviewing (dyad) & Linear SVM  \\
\rowcolor{LightCyan}
\cite{park2013mutual} & CUC, CTM & Smile, posture, VFOA, eye gaze, prosody (voice quality, pitch, energy, spectral stationarity) & Meetings (dyad) & Linear SVM  \\
\cite{xiao2014modeling} & FFM & MFCCs, pitch, turn-taking cues & Motivational interviewing (dyad) & LR, linR  \\
\rowcolor{LightCyan}
\cite{lubold2014acoustic} & Unidirectional CTM & Prosody (intensity, pitch, voice quality, and speaking rate) & Collaborative problem-solving (dyad) & Correlation analysis  \\ 
\cite{xiao2015analyzing} & FFM & Speech rate & Motivational interviewing (dyad) & Linear SVM  \\ 
\rowcolor{LightCyan}
\cite{Muller2018} & LAM, CUC, FSC, CTM & Speaking activity, prosody, hand motion, head orientation, FAU & Meetings (3-4) & SVM w/ RBF \\  
\hline 


\end{longtable}
\normalsize
\end{center}
}

\vspace{-1cm}
\subsection{Interaction Dynamics}
\label{groupdynamics}
Interaction dynamics is a complex phenomenon related to individuals' traits and their roles in the dyad/group. Its analysis includes but is not limited to, the detection of group conversational context, engagement, involvement, group cohesion, empathy, and rapport. Moreover, vocal entrainment is a well-known conversational phenomenon in which the interactants show a synchronization of words and/or speaking style during their conversation. It has a high correlation with engagement, rapport, and even empathy. Through nonverbal behavior analysis, it is possible to detect whether there is a high/low group performance \cite{Zhong_interspeech_2019}, quantify interaction quality \cite{Lahnakoski2020}, or predict group satisfaction level \cite{Lai2018}. Below, we review the corresponding computing studies in depth while their summarization is given in Table~\ref{groupdynamics_table}. \\

\noindent \textbf{\textit{Group Conversational Context.}} Computing studies performed the automatic recognition of group conversations in terms of the context: \emph{brainstorming vs. decision-making} \cite{Jayagopi2012b}, \emph{formal vs. informal} \cite{Matic2014}, \emph{focused vs. unfocused} \cite{Bano2017} and \emph{scenario vs. non-scenario} \cite{Miller2019} interactions. Given that each study focused on different contexts and was tested on different datasets and scenarios, we review them independently by avoiding comparing them in terms of the performance of the used nonverbal cues and computational models. Still, it is noticeable that \textit{i)} every study utilized the speaking activity as a cue, \textit{ii)} three of four works applied SVM for prediction, and \textit{iii)} three of four works used wearable sensors. In \cite{Jayagopi2012b}, brainstorming and decision-making interactions were discriminated by analyzing nonverbal behaviors of individuals (e.g., total speaking length, total speaking turn) and groups (e.g., group speaking length distribution, group speaking turn distribution) when the data was collected with a privacy-sensitive mobile sociometer during meetings. It was found that the fraction of silence, the fraction of overlapped speech, and group speaking length are the most effective features. That study \cite{Jayagopi2012b} differs from others in terms of the sensor utilized, i.e., it did not apply the classical setup of meetings with microphones and cameras. Consequently, participants were able to move freely (notice that authors did not report any sensor error) and were not obliged to sit at a table as majorly happens in several other meeting datasets.
On the other hand, the workplace social interactions were analyzed in \cite{Matic2014} by using smartphones and ACC. Among interpersonal distance, relative body orientation, and speech-based nonverbal features, the speaking activity was the most dominant one. Since the dataset used in \cite{Matic2014} was in the wild and long-durational, the findings of that study are remarkable. The speaking activity was quantified in terms of \textit{a)} participant’s involvement, \textit{b)} the cycles of high/low activity, and \textit{c)} interruptions to discriminate scenario and non-scenario meetings with parallel episodes \cite{Patnaik2008} technique in \cite{Miller2019}. That study was unique due to the computational method it used. Bano et al. \cite{Bano2017} defined focused interactions as the ones the co-present individuals have mutual VFOA, establishing face-to-face engagement and direct conversation. They proposed an online SVM-based classifier to distinguish unfocused from focused interactions in egocentric videos. Visual face track scores, camera motion profiles, and speaking activity showed the best performance when they were fused. That has been the only study detecting in-the-wild indoor/outdoor interactions captured from the first-person perspective and adapting an online learning algorithm. \\

\noindent \textbf{\textit{Engagement, Involvement and Interest Level.}} The computing approaches performing automated detection of engagement, involvement, and interest level were tested on: meetings \cite{bednarik2012gaze,Oertel2013a}, speed-dates \cite{Veenstra2011,Kapcak2019} and poster sessions \cite{kawahara2013estimation}. The spectrum of nonverbal cues used is wide and ranges from proximity \cite{Veenstra2011} to motion \cite{Veenstra2011,Kapcak2019}, gaze analysis \cite{bednarik2012gaze,Oertel2013a,kawahara2013estimation} and acoustic features \cite{wang2013automatic}. Cameras and microphones were dominantly used, instead, a few studies utilized the data sensed by the eye trackers and ACC. The findings of the reviewed research are as follows. Veenstra et al. \cite{Veenstra2011} showed that the video-based features (position, proximity, and body motion) perform better than audio-only systems using standard SVM and k-NN for the engagement detection in speed dates (dyads composed of a male and a female). Xiao et al. \cite{xiao2012multimodal} distinctively relied on Approach-Avoidance (AA) coding that measures the involvement and immediacy and specifically focuses on the salient events in dyadic interactions, which trigger change points in AA code in time. Within this approach, body motion, pose, and vocal energy features were modeled by an SVM, concluding that visual cues are more reliable compared to vocal energy to make decisions on salient AA events. Bednarik et al. \cite{bednarik2012gaze} demonstrated that gaze patterns: fixations, and saccades captured with eye trackers and modeled with an SVM are effective in detecting engagement and its level (i.e., no interest, following, responding, conversing, influencing, governing the discussion) in meeting scenarios. A limitation of that work \cite{bednarik2012gaze} is not performing an ablation study for the window length (fixed to 15 seconds) in which the analysis was being performed. Oertel et al. \cite{Oertel2013a} successfully distinguished the levels of group involvement with GMMs and the gaze patterns (e.g., the fraction of subjects looking at other subjects, mutual gaze, and the maximum fraction of subjects looking at the same target) in a role-playing game.
Kawahara et al. \cite{kawahara2013estimation} showed that the visual cues (gaze occurrence frequency, gaze duration) and particularly the gaze occurrence frequency contribute much more than verbal backchannels (e.g., "yeah", "ok") to estimate the interest level of the audience in poster sessions where people stand. Wang et al. \cite{wang2013automatic} found out that acoustic features (e.g., low-level acoustic energy, speech duration, voice quality features, MFCCs) represented for each utterance dominate all lexical features to automatically detect the level of interest. The conclusions of that study are debatable since the language of the used datasets was German while the model used for prosodic analysis was trained in English.
To automatically detect the attraction (notice that this is relevant to engagement) in speed dates using movement features that were captured by single body-worn ACC, Kapcak et al. \cite{Kapcak2019} first time proposed motion convergence (measuring if two people’s behavior style is symmetric) and proved its effectiveness with LR. The research efforts so far discussed in this subsection relied on annotations based on the perception of third persons. Even though studies significantly differ from each other in terms of the datasets used for model evaluation, one can still conclude that: \textit{i)} overall, visual cues performed better than acoustic features, \textit{ii)} acoustic features in some cases were preferable to lexical features, and \textit{iii)} gaze activity and body motion were among the most effective cues to detect the phenomena of interest. \\

\noindent \textbf{\textit{Group Performance.}} Automatic detection of group performance is a task investigated mostly in small groups (3-4 people) in which people make a decision regarding a (pre-defined) task \cite{Nihei2014,Avci2014,Avci2016,Zhong_interspeech_2019,Go2019,Kubasova2019,Lin2020}. That topic is important since an intelligent detection system can provide information aimed at enhancing the performance and efficiency of a group and its members. This can be used by managers to better understand the dynamics of their groups, improve the productivity of meetings, and prevent organizations from wasting time and money \cite{Kubasova2019}. Meeting environments constituted to capture relevant data, were equipped with microphones (ARM, LAM, HSM) and/or cameras (WC). The findings, technical novelties, and limitations of the reviewed works are summarized as follows. Nihei et al. \cite{Nihei2014} showed that the influential statements in group discussions, which affect the discussion flow and are highly related to group performance, can be automatically predicted in terms of prosody, VFOA, and head motion features modeled with SVM, RF, and LR. By modeling audio-visual behavioral features with multivariate binary IM, \cite{Avci2014} demonstrated that different group performance clusters have different interaction types. The same authors extended their feature space in \cite{Avci2016} by involving self-reported features regarding personality, perception-related, and hierarchy in the group and showed that self-reported features were important while the group-looking cues and the influence cues (i.e., the confidence score IM model produces) were major predictors for group performance while IM outperformed HMM, on average. Linguistic features leveraged the performance of acoustic features (MFCCs, associated delta features, jitter, shimmer, PCM loudness, F0 envelope, F0 contour, voicing probability, and log power of Mel-frequency bands) to predict group performance in task-based interactions in \cite{Murray2018}. The same study also presented an interesting data augmentation strategy. First, a model with the target dataset was trained and used to annotate an auxiliary dataset. Then, a new model for the prediction of group performance was trained with the target and the auxiliary datasets. That final model was used to make predictions on the test split of the target dataset, resulting in improved performance compared to not applying data augmentation. We believe that such an approach can be adapted for other traits' automatic detection to increase the model's accuracy. All aforementioned efforts were limited to being tested on relatively small-scale datasets. Instead, Zhong et al. \cite{Zhong_interspeech_2019} proposed a novel network composed of DNN and BLSTM with an attention mechanism, which models the vocal behaviors and personality to predict the group performances during collaborative problem-solving tasks, tested on a large data collection. Including personality attributes showed 14\% performance improvement while vocal behaviors were more significant in the high-performing groups than the low-performing groups \cite{Zhong_interspeech_2019}. Lin et al. \cite{Lin2020} utilized conversation dynamics as the graph to aggregate group members’ speech and lexical behaviors within a deep model surpassing the SOTA results of several datasets. The experimental analysis in \cite{Lin2020} was extensive compared to its counterparts and as being validated on three diverse datasets while performing the best out of all, it provided more robustness. It is notable that nonverbal features tend to be used together with lexical features for the detection of group performance. This situation is different from other phenomena (reviewed in previous subsections) that were detected only with nonverbal cues. Overall, using the features representing personality in addition to the acoustic features improved the detection performance. The only work analyzing the long-term interactions collected from wearable sensors in this domain is \cite{Yanxia2018}, which analyzed the continuous tracks of six persons for four months in a mission with SB. That study showed that considering behaviors of individuals at different temporal resolutions contributes.
\\
\\
\textbf{\textit{Group Satisfaction.}} Estimating group satisfaction automatically is crucial for developing strategies for computer-aided decision-making and human-robot interaction while also being important to understanding the cognitive state of individuals. However, this topic (at least after 2010) has not been studied much within co-located human-human social interaction analysis. Lai et al. \cite{Lai2018} performed regression analysis showing that combining acoustic, lexical, and turn-taking features improves performance while prosodic features are good at attention satisfaction prediction and voice quality is more informative for predicting information overload. The weak points of that paper \cite{Lai2018} are; not validating their findings on other datasets and not proposing novel computational methods or nonverbal cues.
On the other hand, a recent short paper \cite{johnson2021clustering} presented the estimation of group satisfaction by clustering the features composed of pronoun usage and coordination (extracted by Convokit \cite{chang2020convokit}, a measure which reflects the change in speaking patterns to become more linguistically similar to others within groups). That approach is innovative because of its concept and also relies on clustering rather than predicting a particular outcome of interest. However, its experimental analysis is limited to a single dataset. It is noticeable that both studies focused on speaking activity and there exists no work investigating the effectiveness of several other cues (especially vision-based) for the automatic group satisfaction estimation.
\\
\\
\textbf{\textit{Quality of Interactions.}}
There exist studies that addressed the automated estimation of the quality of co-located human-human social interactions using physiological signals such as \cite{chaspari2015quantifying,chaspari2017exploring} which processed electrodermal activity. On the other hand, it is well-known that physiological activity is correlated with nonverbal behaviors and vice versa. Though, surprisingly, there exists only one work \cite{Lahnakoski2020}, which implemented detection of the quality of interactions through nonverbal behavior analysis. Lahnakoski et al. \cite{Lahnakoski2020} considered movement synchrony in addition to several other nonverbal features modeled with NB and SVM. It was exhibited that \textit{a)} proxemic behaviors best predict the quality of the interactions rather than interpersonal movement synchrony, \textit{b)} increased distance between participants predicts lower enjoyment, \textit{c)} increased joint orientation towards each other during cooperation is correlated with increased effort, and \textit{d)} the interpersonal distance is not informative to estimate the quality of the interaction. These results were obtained by processing the data captured with Kinect and IMU sensors. The related literature lacks the use of acoustic features, multimodal cues, and deep models while all these options have the potential to boost the detection performance.
\\
\\
\textbf{\textit{Group Cohesion.}} Research in social and organizational psychology has shown that good cohesion between group members is correlated with team effectiveness, productivity, and performance. Cohesion is split into two: social and task cohesion. Social cohesion refers to the attractiveness (likeliness) of group members towards each other \cite{Nanninga2017}. This can be extracted by asking questions such as whether the teammates appeared to be involved/engaged in the discussion or have a good rapport. Task cohesion, on the other hand, is related to how much a group is reaching its goals in a shared way. Therefore, one can measure task cohesion by asking questions such as whether the group members share responsibilities and goals or whether the teammates are collaborative \cite{Hayley2010}. Meetings have been suitable environments to perform automatic group cohesion detection \cite{Hayley2010,Nanninga2017,Zhang2018}. Unlike other works, \cite{Sharma2019} analyzed diverse interactions by processing YouTube videos containing events such as interviews, festivals, and parties. The major findings and novelties are discussed as follows.   
Hung et al. \cite{Hayley2010} estimated the high/low group cohesion using audio (e.g., speaking energy, turn duration, speaking time), video and audiovisual cues (e.g., upper body motion during the overlapping speech, motion when not speaking) with NB and SVM. The best-performing feature was found as a turn-taking feature which is the accumulation of the total pause duration between each individual's turns. Instead, mimicry features in terms of speaking rate and frequency (defined as how well subsequent samples from subjects’ audio fit the learned distributions from the other participants) performed better than the turn-taking features of \cite{Hayley2010} when they were modeled with a Gaussian NB, moreover, for social cohesion, the performance of the turn-taking and mimicry features were found comparable \cite{Nanninga2017}. Differently, aggregating the behavior patterns extracted from individuals' SB data with the group-level interactions improved the effectiveness of assessing group cohesion in meetings composed of up to six persons in \cite{Zhang2018}. Sharma et al. \cite{Sharma2019} proposed a two-head (one for visual data and the other for acoustic data) DNN architecture composed of CNNs and LSTMs. To obtain visual and acoustics embeddings, the Inception V3 model pre-trained on GAF-cohesion database \cite{ghosh2019predicting} and OpenSMILE \cite{Eyben2013} tool were used, respectively. That work was unique as it is based on deep learning for group cohesion detection tasks. Overall, all works included acoustic features, implying that they are effective for such an automated task, and the performance could further be boosted by the integration of head and body movement cues \cite{Hayley2010,Zhang2018}. \\

\noindent \textbf{\textit{Vocal Entrainment.}} 
Interaction synchrony in human-human conversations can occur naturally and spontaneously. Vocal entrainment is the phenomenon in which the interactants show a synchronization of speaking style during their conversation. The automatic detection of such a phenomenon was investigated in dyads when both participants were wearing ARM \cite{lee2010quantification,lee2010quantification,georgiou2011behavioral}. Reviewed works focused on discovering the effective acoustic features with statistical modeling \cite{lee2010quantification}, in terms of correlations \cite{lee2012using} and with SVM \cite{georgiou2011behavioral}. There exists no work presenting a data-driven approach and/or processing the raw audio data without requiring an additional feature extraction step. Lee et al. \cite{lee2010quantification} exhibited the success of turn-wise entrainment measures obtained from prosodic cues, specifically pitch and energy to understand the overall attitude of the interacting partners in problem-solving interactions. Georgiou et al. \cite{georgiou2011behavioral} built an SVM classifier with the features of \cite{lee2010quantification} to analyze distressed dyadic interactions in terms of vocal entrainment and concluded that including video-based cues or body sensors' data can improve the task. Furthermore, KCCA in \cite{lee2012using} showed that there exists a statistically significant relationship between vocal entrainment and withdrawal, which were explained based on the variations in the overall vocal engagement level in the problem-solving interactions. The experiments in all studies were limited to a single dataset targeting a very specific scenario: married couples' interactions.\\

\noindent \textbf{\textit{Rapport and Empathy.}} 
Empathy is an important skill in developing rapport with a person \cite{xiao2015analyzing}. It means close and harmonious relationships where the interaction patterns are synchronized while attention, positivity, and coordination are necessary \cite{Muller2018}. Rapport is a complex social behavior, and low rapport can result in interpersonal conflicts and decreased collaboration. The data was captured with microphones (LAM, FFM, CTM) and cameras (CUC, FSC) to perform automatic detection of rapport and empathy during the meeting of 2-persons \cite{Hagad2011,park2012already,park2013mutual}, 3-4 participants \cite{Muller2018}, political debates of up to 5-persons \cite{song2012multimodal}, motivational interviews of 2-persons \cite{xiao2013modeling,xiao2014modeling,xiao2015analyzing} and collaborative problem-solving in dyads \cite{lubold2014acoustic}. This implies that related works used different datasets, which does not allow us to make a performance comparison among them. Therefore, we discuss the findings with respect to the effectiveness of the nonverbal cues, as follows.
Posture mirroring showed a powerful performance in detecting the rapport with NB, SVM, and MLP in \cite{Hagad2011}. Muller et al. \cite{Muller2018} analyzed facial expressions, hand motion, gaze, the speaker turns, and prosody to detect low rapport, where facial features performed on average the best and incorporating the participant's personalities (which is a particular novelty of this work) as a prior knowledge contributed positively. 
MFCCs, pitch, and turn-taking features alone and combined were found significantly and highly correlated with empathy, respectively \cite{xiao2013modeling}. High pitch and energy were found negatively correlated with empathy \cite{xiao2014modeling}.
In \cite{xiao2015analyzing}, the average absolute difference of turn-level speech rates between two people, silence duration, and other statistics of speech was exhibited as correlated with empathy. Furthermore, for automatically predicting high/low empathy with SVMs, speech rate cues were found effective, meaning that vocal entrainment contributes to empathy modeling \cite{xiao2015analyzing}. 
On the other hand, \cite{lubold2014acoustic} demonstrated the correlation between entrainment and rapport. In detail, the speakers appeared to entrain primarily by matching their prosody on a turn-by-turn basis and the pitch was the most significant prosodic feature people entrain on when rapport was present. 
What differs \cite{lubold2014acoustic} from \cite{xiao2013modeling,xiao2014modeling,xiao2015analyzing} is the scenario relied on: motivational interviewing vs. collaborative problem-solving. 
The recognition of the agreement and disagreements is also related to empathy and rapport. 
Multi-view HCRF was presented to be effective in learning interactions in terms of the audio-visual cues (head nod, head shake, forefinger raise, forefinger raise-like, forefinger wagging, hand wag, hands scissor, shoulder shrug) and acoustic features (fundamental frequency and energy) when KCCA was used to capture the nonlinear hidden dynamics in \cite{song2012multimodal}. Using head nodding, head shaking, head tilting, eye gaze, smiling, and self-touching together was suggested in \cite{park2012already} to predict the acceptances or rejections of proposals in a dyadic negotiation.
The same authors \cite{park2013mutual} detected that symmetric smiles, symmetric postures, mutual gaze, symmetric and asymmetric head orientation, asymmetric eye gaze, asymmetric voice pitch, asymmetric voice quality, and asymmetric spectral stationarity are useful in predicting the respondent reactions. In conclusion, prosody, speaking, head, and body activities are competent for rapport and empathy detection. Future research can consider applying audio-visual deep learning methods.
{\small\tabcolsep=3pt  
\begin{center}

\footnotesize
\begin{longtable}{|p{2.3cm}|p{3.2cm}|p{1.5cm}|p{4.5cm}|p{2cm}|p{0.5cm}|}

\caption{Related datasets chronologically summarized in terms of scenarios, group size (GS), the total number of participants (\#Par.), annotations, sensors, and public availability (Pub). $S$ is for self-annotations, $E$ refers to the annotation performed by external observers, and $A$ represents the automatic methods. ``cov.'' and ``int.'' are used for conversations and interactions, respectively.} \label{table:dataset} \\

\hline \multicolumn{1}{|c|}{\textbf{Ref.}} & \multicolumn{1}{c|}{\textbf{Scenarios}} &
\multicolumn{1}{c|}{\textbf{GS / \# Par.}} &
\multicolumn{1}{c|}{\textbf{Annotations}} &
\multicolumn{1}{c|}{\textbf{Sensors}} &
\multicolumn{1}{c|}{\textbf{Pub}} \\ \hline 
\endfirsthead

\multicolumn{6}{c}%
{{\bfseries \tablename\ \thetable{} -- continued from previous page}} \\
\hline \multicolumn{1}{|c|}{\textbf{Ref.}} & \multicolumn{1}{c|}{\textbf{Scenarios}} &
\multicolumn{1}{c|}{\textbf{GS / \# Par.}} &
\multicolumn{1}{c|}{\textbf{Annotations}} &
\multicolumn{1}{c|}{\textbf{Sensors}} &
\multicolumn{1}{c|}{\textbf{Pub}} \\ \hline 

\endhead

\hline \multicolumn{6}{|r|}{{Continued on next page}} \\ \hline
\endfoot

\hline \hline
\endlastfoot

\rowcolor{LightCyan}
Idiap Wolf \cite{Wolf2010} & Game-play, sitting conv. &  8-12 / 36 &  Roles (E), speaking activity (E) & HSM, ARM, WC & $\checkmark$ \\

Cocktail-party \cite{Zen2010} & Indoor free-standing conv.  &  6-7 / 13 & Extroversion ($S$), neuroticism ($S$) & FFC, PTZ & $\checkmark$ \\

\rowcolor{LightCyan}
\cite{rozgic2010new} & Conv. on conflictual topics w/ role-play  &  Dyad & Subject-int. level, approach-avoidance ($E$) & LAM, ARM, Vicon MCS, point grey camera array &  \\

Coffee Break \cite{Cristani2011} & Indoor free-standing conv. & 6-14 & Social groups ($E$) & Fish-eye camera & $\checkmark$ \\

\rowcolor{LightCyan}
\cite{Feese2011b} & Role-play meeting & 3 / 126 & Actions (e.g., face-touch) ($E$), leadership style ($E$) & IMU & \\ 

\cite{Chen2011} & Indoor daily int. in a research group & 3-5 / 5 & Social groups ($A$), participants' location ($A$), head orientation ($A$) & RGB cameras, RFID &  \\ 

\rowcolor{LightCyan}
Twenty-Question Game \cite{Kalimeri2011} & Brainstorming and problem-solving games & 4 / 52 & Dominance ($S$), group performance ($E$) & SB (Prox., ACC, microphone) &  \\ 

\cite{Srivastava2012} & 3907 movie clips & - / 50 & Emotions ($A$), personality traits ($A$) & DMC &  \\ 

\rowcolor{LightCyan}
SocioMetric Badges \cite{Lepri2012b} & Daily activities in a research institute for 6 weeks & 53 & Personality traits ($S$), affective states ($S$), perceived productivity and relations ($S$) & SB (microphone, infrared beam detector, Bluetooth detector, and ACC) &  \\  

ELEA \cite{SanchezCortes2012} & Meeting (survival task) & 3-4 / 148 & Personality traits ($S$), leadership ($S$, $E$), competence ($S$), likeness ($S$), dominance ($S$) & LAM, ARM, WC & $\checkmark$ \\

\rowcolor{LightCyan}
First-Person Social Int. \cite{Fathi2012} & Day-long videos in a theme park &  8 & Social int. type (e.g., dialogue, discussion) ($E$), activities (e.g., walking) ($E$) & GoPro & $\checkmark$ \\

D64 \cite{Oertel2012} & Free-sitting conv. & 4-5 & Overall group excitement ($E$), pairwise distance ($E$) & HSM, LAM, MoCAP & $\checkmark$ \\

\rowcolor{LightCyan}
KTH Wolf \cite{Oertel2013} & Role-playing game, sitting participants & 8 & Eye gaze ($E$), voice activity ($E$), group involvement ($E$) & CUC & $\checkmark$ \\

YouTube Social Roles \cite{Ramanathan2013} & YouTube event videos (e.g., birthday party) & 964 videos & Event names ($E$), social roles ($E$) & DMC & $\checkmark$ \\

\rowcolor{LightCyan}
SONVB \cite{Nguyen2014} & Real job interviews & 3 / 62 & Hirability ($E$), personality traits ($S$), intelligence ($S$), communication and persuasion skills ($S$) & ARM, CUC &  \\ 

EGO-GROUP \cite{Alletto2014} & Indoor / outdoor int. & 13 clips & Group identifier ($E$) & Narrative camera & $\checkmark$ \\

\rowcolor{LightCyan}
MATRICS \cite{Nihei2014} & Role-based meetings & 4 / 40 & Personality traits ($S$), (not)influential statements ($E$) & MoCAP, ACC, Kinect, ET, FSC, WC, HSM & \\ 

KTH-Idiap \newline Interview. \cite{Oertel2014} & Role-play group interviews & 4 / 20 & Eye gaze ($E$), voice activity ($E$), personality traits ($S$), interest-level/performance/potential related questionnaire ($E$), performance ($E$) & Kinect, GoPro, CTM & $\checkmark$ \\

\rowcolor{LightCyan}
MAHNOB \cite{MAHNOB2015} & Dyadic discussions and negotiations & 2 / 60 & Mimicry of head gestures, hand gestures, body movement and facial expressions ($E$), reflective mimicry ($E$) & CUC, HSM, FFM & $\checkmark$ \\

SALSA \cite{AlamedaPineda2015} & Indoor free-standing conv. & 18 & Personality traits ($S$), position ($E$), head/body orientation ($E$), F-formation ($A$) & FSC, SB (microphone, infrared beam detector, Bluetooth detector, and ACC) & $\checkmark$ \\

\rowcolor{LightCyan}
PAVIS Leadership \cite{Beyan2016,Beyan2020ICMI} & Meeting (survival task) & 4 / 64 & Leadership ($S$, $E$), designate leaders ($S$), leadership style ($E$), voice activity ($E$), face-touching behavior ($E$) & LAM, CUC & $\checkmark$ \\

MobileSSI \cite{Flutura2016} & Free discussions in the pub & 3 & Voice activity ($E$), laughter ($E$) & LAM, smartphone (ACC) & \\ 

\rowcolor{LightCyan}
\cite{Terven2016} & Meeting, topic-defined conv. & 2 / 48 & Head nodding ($E$), conv. usefulness ($E$), sustained concentration ($E$), competence ($E$), satisfaction level ($E$) & Smart glasses, CUC &  \\ 

\cite{Nanninga2017} & Business meeting & 3-8 / 107 & Cohesion ($E$) & ARM, HSM, HD cameras, ACC &  \\ 

\rowcolor{LightCyan}
\cite{Ohnishi2017} & Storytelling events & 14 & Interest level ($E$), observed action ($E$) & ACC, Gyro, HD camera &  \\ 

MatchNMingle \cite{MatchNMingle} & Indoor free-standing conv. and speed dates & 8-2 / 94 & Personality traits ($S$), self-control scale ($S$), sociosexual orientation inventory ($S$), social cues ($E$), social cautions ($E$), F-formations ($E$) & Triaxial ACC, Prox., GoPro, TVC & $\checkmark$ \\ 

\rowcolor{LightCyan}
MULTISIMO \cite{koutsombogera2018} & Meeting (solving a quiz) & 3 / 49 & Personality traits ($S$, $E$), experience ($S$), speaking activity ($A$), dominance ($E$), transcripts ($E$), turn-taking ($A$), emotions ($A$) & CUC, C360, HSM, ODM, Kinect & $\checkmark$ \\

AMIGOS \cite{Correa2018} & Watching videos & 4 / 20 & Personality traits ($S$), mood ($S$), valence ($S$, $E$), arousal ($S$, $E$), dominance ($S$), liking ($S$), familiarity ($S$), emotions ($S$) & EEG, ECG and skin conductance sensors, CUC, Kinect & $\checkmark$ \\

\rowcolor{LightCyan}
MPII Group Interaction \cite{Muller2018} & Meeting, topic-defined conv. &  3-4 / 78 & Rapport ($E$), leadership ($S$), dominance ($S$), competence ($S$), liking ($S$), personality traits ($S$) & LAM, CUC, FSC, CTM & $\checkmark$ \\

Focused Interaction \cite{Bano2018} & Daily-life indoor/outdoor int. & 19 sessions & (No)focused interaction ($E$), voice activity ($A$) & GoPro, ACC, GYRO, MAG, smartphone (GPS) & $\checkmark$ \\ 

\rowcolor{LightCyan}
UGI \cite{Bhattacharya2018} & Meeting (survival tasks) & 4 / 36 & Perceived leadership ($E$), perceived contribution ($E$) & Time-of-flight sensors, Kinect, LAM & $\checkmark$ \\

SRIV \cite{Jinna2018} & Movies and TV dramas & 69 movies, 3124 videos & Subjective and objective social relations ($E$) & DMC &  \\ 

\rowcolor{LightCyan}
TeamSense \cite{Zhang2018} & Four-month simulation of a space exploration mission &  6 & Roles ($E$), affective states ($S$), team cohesiveness ($S$) & SB (microphone, infrared beam detector, Bluetooth detector, and ACC) & \\ 

Panoptic \cite{Joo2019} & Standing game-play & 3-12 & Roles ($E$), speaking activity ($E$) & HD and VGA cameras, Kinect & $\checkmark$ \\

\rowcolor{LightCyan}
GAP \cite{Braley2018} & Meetings (survival task) & 2-3 / 37 & Group satisfaction ($E$), group decision-making ($E$), dialogue acts ($E$), sentiment ($E$), transcription ($E$) & WC, handy cam & $\checkmark$ \\

VGAF \cite{Sharma2019} & Web videos & 1004 videos & Group-level emotion ($E$), group-level cohesion ($E$) & DMC & $\checkmark$ \\

\rowcolor{LightCyan}
EgoSocialRelation \cite{Aghaei2016,Aimar2019} & Daily-life indoor and outdoor int. & 8 first-person / 100 & Sequence annotation ($E$), face locations ($E$),
social int. state ($E$), social relations ($E$) & Narrative camera & $\checkmark$ \\

\cite{Katevas2019} & Social networking event & 22 & Social groups ($E$) & Smartphone (iBeacon Prox., ACC, Gravity, rotation rate), HD cameras & \\ 

\rowcolor{LightCyan}
ViSR \cite{Liu_2019_CVPR} & Movies & 8000 clips & Social relations ($E$) & DMC & $\checkmark$ \\

\cite{Echeverria2019} & Healthcare simulations & 4 / 9 & Roles ($E$), collaboration translucence ($E$) & Skin conductance sensor, ACC, ARM, HD camera & \\ 

\rowcolor{LightCyan}
Resistance Game \cite{Chongyang2019} & Game-play & 5-8 / 233 & Roles ($E$), dominance ($E$) & HD video camera & $\checkmark$ \\

NTHULP \cite{Zhong_interspeech_2019} & Role-based meetings & 3 / 194 & Roles ($E$), personality traits ($S$), group performance outcome ($E$) & CUC, LAM &  \\ 

\rowcolor{LightCyan}
\cite{Othman2019} & Game-play, standing conv. & 4 / 96 & Transcriptions ($E$), text binary content parameters ($E$), group contribution ($E$) & CUC, HSM &  \\ 

GAME-ON \cite{Maman2020} & Game-play, standing conv. & 3 / 51 & Cohesion ($S$), leadership ($S$), emotional state ($S$), warmth and competences ($S$) & IMU (ACC, GYRO, MAG), infrared, RGB cameras, HSM & $\checkmark$ \\

\rowcolor{LightCyan}
WoNoWa \cite{Biancardi2020} & Indoor free-standing conv. performing a pre-defined task & 3 / 45 & Voice activity ($A,E$), quantity of motion ($A$), head position and rotation ($A$), chronemics ($E$), proxemics ($E$), kinesics ($E$), leadership ($S$), degree of acquaintance ($S$), warmth and competence ($E$), perceived transactive memory system ($E$) & HD handy recorder, HSM & $\checkmark$ \\ 

\cite{Onishi2020} & Topic-predefined meeting, sitting participants & 2 / 34 & Dialogue data ($E$), praising skills ($E$) & CUC, HSM &  \\ 

\rowcolor{LightCyan}
\cite{Lahnakoski2020} & Topic pre-defined conv., cooperative and competitive game-play &  2 / 18 & gaze ($E$), head location/orientation ($E$), enjoyment ($E$), choice ($E$), felt pressure ($E$), effort ($E$), personality traits ($E$) & Kinect, IMU &  \\ 

\cite{Zhang_2020_WACV} & Meeting (survival task) & 3-4 / 48 & Personality traits ($S$), contribution level ($E$), leadership ($E$) & CUC, Kinect &  \\ 

\rowcolor{LightCyan}
VYAKTITV \cite{khan2020vyaktitv} & Daily-life conv.  & 2 / 25 & Personality traits ($S$) & CUC, FFM &   \\

UDIVA \cite{Palmero_2021_WACV} & Competitive and collaborative tasks & 2 / 147 & personality traits ($S$, $E$), sociodemographics ($S$), mood ($S$), fatigue ($S$), social relation ($S$) & LAM, CUC, ODM, egocentric CAM, wearable heart rate monitor & $\checkmark$ \\

\rowcolor{LightCyan}
MUMBAI \cite{doyran2021mumbai} & Board-game & 4 / 58 & Personality ($S$), game-playing experience ($S$), emotions ($E$), int. types, e.g., cooperative ($E$) & CUC & $\checkmark$ \\

\end{longtable}
\normalsize
\end{center}
}
\vspace{-1cm}
\section{Datasets}
\label{datasets}
The reviewed works in Sec.~\ref{IntAnalysis} have been conducted using datasets that typically differ from each other in terms of a) the scenarios (e.g., interviewing, brainstorming in a meeting, poster presentation, game playing), b) interaction settings (e.g., the number of people involves, indoor/outdoor, standing/sitting), c) annotations (e.g., self-assessments of participants, the assessment made by observers), and d) the sensing technology used.
In Table \ref{table:dataset}, we present a summary of co-located human-human social interaction datasets (listed chronologically), supplying annotations to detect social and psychological phenomena. As seen, the majority of the efforts were given to small group interactions, typically composed of four persons as well as dyads. When earlier datasets were more based on role-play interactions, recently, several datasets were collected in less unconstrained environments (e.g., poster sessions, speed dates, pub discussions). There exist attempts to curate datasets from YouTube videos, movies, or from a first-person (egocentric) perspective. More discussions on datasets and our proposals regarding data collection can be seen in Sec.~\ref{future_dataset}.

\section{Proposed Future Research Directions}
\label{future}
We propose several future research directions in the context of the automatic detection of social and psychological phenomena in co-located human-human social interactions through nonverbal behavior analysis. They are mostly motivated by the limitations of the works reviewed in Sec.~\ref{IntAnalysis} and Sec.~\ref{datasets}. The content is divided into three: a) Artificial Intelligence (Sec.~\ref{future_ML}), b) dataset collection (Sec.~\ref{future_dataset}), and c) privacy-preserving social interaction analysis (Sec.~\ref{future_Privacy}). 
\subsection{Artificial Intelligence} 
\label{future_ML}
This paper shows that the progress in the field of co-located human-human social interaction analysis is considerable while various AI concepts have been well integrated. The automatic detection of several social and psychological phenomena was addressed by (data-driven) \textbf{deep learning methodologies}, which show better performance compared to machine learning methods with hand-crafted features. Still, the deep models have not been applied for the detection of several phenomena. These are hirability detection, group conversational context classification, group satisfaction detection, estimating the quality of social interactions, vocal entrainment detection, and rapport/empathy detection. On the other hand, the adoption of deep learning has required the need for large-scale datasets. However, the labeling cost is presumably a limiting factor. In this respect, utilizing \textbf{domain adaptation} methods could be considered to label the new datasets. To date, only a few attempts have shown to perform analyses across datasets \cite{Aran2013b,Muller2019}, although there have been several datasets aiming to detect the same or related social phenomena, collected with very similar sensor setups and scenarios while using the same nonverbal signals. 
An alternative approach can be relying on \textbf{unsupervised pre-training} where the labels are not needed to learn the feature embeddings, and only a few annotated data can be used to evaluate the performance of the methods. Unsupervised pre-training was recently tested on several domains, e.g., HAR \cite{paoletti2021unsupervised},  multimodal emotion recognition \cite{Franceschini2022,koromilas2021unsupervised} and sentiment analysis \cite{koromilas2021unsupervised}. Notice that, the last two domains both used nonverbal signals to make decisions and the effectiveness of unsupervised pre-training in \cite{paoletti2021unsupervised,Franceschini2022,koromilas2021unsupervised} was on par with or even better than the several fully-supervised SOTA. Unfortunately, unsupervised pre-training has not been yet integrated into the automated detection of social and psychological phenomena in co-located human-human social interactions.

AI models have been used to make important decisions regarding diverse social and psychological phenomena, and it is important to understand how models make predictions. This issue becomes even more important considering the fact that the SSP domain is very multi-disciplinary.
To address this, \textbf{explainability analysis} (refers to the details and reasons for the prediction of a model that should be clear and easy to understand) should be performed. Some methodologies, like SVMs, MLPs, CNNs, and RNNs, lack \textbf{algorithmic transparency} and \textbf{interpretability}. Therefore, utilizing visualization techniques on images (such as heatmaps, attribution), using attention mechanisms (for instance Grad-CAM \cite{Selvaraju2017}), or applying feature visualization at different levels of a certain network are good alternatives to improve the trained models' explainability. Likewise, SHAP (SHapley Additive exPlanations) \cite{NIPS2017_8a20a862} allows us to understand how each feature impacts the model’s outcome, and to extract the salient segments and features driving a model to make a decision. However, we noticed that the related art has so far stayed behind in addressing the explainability and interpretability of their proposed methods as only a few studies, e.g., \cite{Beyan2019,Beyan2020ICMI} have adapted the aforementioned or similar tools, allowing to achieve a better level of communication across experts of different disciplines. It is also essential to discuss the challenges in \textbf{transferring human social interaction research into daily life}. Many real-world challenges, such as missing sensors in the inference time and domain-shift problems, have not been investigated by the reviewed works. The integration of \textbf{online learning methods} and \textbf{adaptive models} could be considered with this respect. The authors also speculate that once less restrictive (i.e., in-the-wild) datasets are collected, the new methodologies trained on them can be more resilient to handle such real-life challenges.
\subsection{Dataset Collection} 
\label{future_dataset}
Sec.~\ref{datasets} shows that major efforts were performed in collecting and annotating related datasets. However, not all these resources are \textbf{publicly available} (see Table \ref{table:dataset} for details), and it is still relatively challenging to access data suitable for the experimental activity. This is a very important problem that should be addressed as it does not allow researchers to perform a comparative study. One possible reason for not supplying publicly available datasets is that collection and dissemination of human data is an activity that must respect rigorous ethical standards ensuring the protection of individuals' privacy. The most suitable solution is the involvement of ethical committees operating in the institutions that collect the data and the definition of standard protocols to be respected by the whole research community.

The size of the existing datasets tends to be limited compared to other research areas. As an example, if we consider the Affective Computing domain, one can see much larger datasets (e.g., CMU-MOSEI \cite{zadeh2018multimodal} having 23500 video clips, AffectNet \cite{mollahosseini2017affectnet} having millions of annotated images). Whereas the reviewed studies used datasets having on the average up to 50 participants and the number of videos reaches up to 3000 even though collected through crowd-sourcing (see Table \ref{table:dataset}).
This is a problem for the application of AI methods such as deep learning that typically requires \textbf{large amounts of data} to be effectively trained. Furthermore, models trained on limited material might not generalize well, and thus perform as desired in some conditions but not in others. In this respect, creating large-scale benchmarks such as ImageNet \cite{imageNet} (a large visual database designed for use in visual object recognition) or ActivityNet \cite{caba2015activitynet} (benchmark aims at covering a wide range of complex human activities) would contribute to the field positively. The creation of such \textbf{benchmarks} would allow testing and more importantly comparing different methodologies to extract and/or learn in a data-driven way the nonverbal behaviors, which are further used for the detection of various social and psychological phenomena. Another important thing to consider for collecting data is to rely on \textbf{real-world scenarios}. It is also interesting to collect \textbf{long duration data} where participants are involved in multiple tasks and diverse environments and interact with various people for several days or weeks. Interpersonal behavior differences especially the ones arising due to \textbf{ethnic diversity} are a challenge for social behavior analysis. Therefore, future work can focus on collecting datasets including subjects with diverse cultural backgrounds as well.

The experiments were often made by use of psychometric questionnaires. In some cases, they are self-assessments provided by the participants whereas in others they are assessments about the participants provided by external observers. In both cases, the questionnaire outcomes are used as labels for the evaluation of the approaches. Therefore, in the case of assessments made by multiple observers, it would be important to analyze \textbf{the reliability of the judgments}, i.e., whether the different observers actually agree beyond the chance level. The aforementioned reliability analysis is still not in the common practice of reviewed datasets. For instance, studies \cite{Favaretto2019,Zhang_2020_WACV,Chongyang2019,Jinna2018,Jinna2019,Matic2014,Tung2012,Echeverria2019,Bhattacharya2018,Solera2015,Suzuki2013,Hagad2011,Bhattacharya2019,Tung2014,koutsombogera2018,Onishi2020,Kumano2015,Bano2018,Alletto2014,Ramanathan2013,Alletto2015,Yasuyuki2010,Nihei2014,Bano2017} neither discussed the reliability of the annotations nor showed the trustfulness of the ground-truth data by applying quantitative inter-rater agreement analysis.
\subsection{Privacy-preserving Social Interaction Analysis} 
\label{future_Privacy}
Recording and storing signals from sensors might violate the \textbf{privacy} of a person especially when consent has not been explicitly obtained. Besides, preserving the privacy of individuals is essential once the research is transferred into real-world applications. There exist related works considering the usage of privacy-sensitive sensors such as \cite{Jayagopi2012b} utilizing mobile sociometer. However, our review shows that there is no study performing \textbf{privacy-preserving nonverbal signal processing}. Haider and Luz \cite{Haider2019} presented a low-cost system that was used to extract audio features while the actual spoken content was protected. Sajadmanesh and Gatica-Perez \cite{Sajadmanesh2020} developed a privacy-preserving GCNN learning algorithm based on Local Differential Privacy, which is important to apply for the problems where graph nodes contain sensitive features that need to be kept private while they could be beneficial for a central server for training the GCNN. But, neither these studies \cite{Haider2019,Sajadmanesh2020} nor others have been integrated into the detection of social and psychological phenomena. Future research could focus on answering the questions: \textit{a)} what are the privacy-preserving nonverbal signal representations? (furthermore, is not using features extracted from appearance enough, is using the audio features having low linguistic information enough?), \textit{b)} What are the privacy-preserving sensors, \textit{c)} How to prove that one approach is more privacy-sensitive than another? (for instance, can lower intelligibility be used for that?), and \textit{d)} Can we preserve the performance when using more privacy-sensitive features?

\section{Conclusions} 
\label{conclusion}
This survey paper has gathered together the research efforts shown regarding the automatic analysis of nonverbal cues displayed in co-located human-human social interactions. We assembled peer-reviewed papers since 2010, which proposed methodologies to automatically analyze and estimate social and psychological phenomena including social traits, social roles, social relations, and several others related to interaction dynamics. Our article significantly differs from the earlier surveys due to its broad coverage. Additionally, we provide a review of related datasets. We also propose several future research directions, not yet being addressed by the SSP-oriented HBU studies.
Our generalizable key observations are outlined as follows.
        \textit{i)} The studies investigating the same phenomena diverge from each other in terms of the computational methodology they proposed, the nonverbal cues they used, the interaction environment and/or scenarios they tackled, or the sensors they utilized. Addressing one or a couple of such items in a different way from the prior art made the more recent novel/original.
    \textit{ii)} Several reviewed papers showed that using multimodal features improves inference performance.
    \textit{iii)} The applied computational methods are very varied. Some studies presented novel pipelines rather than only applying mainstream methods. In general, the deep models improved the performance significantly for the automatic detection of some phenomena. However, SVM was the classifier that was utilized the most.
    \textit{iv)} Automatic detection of personality traits was studied the most out of all social phenomena.
    \textit{v)} The speaking activity was the most preferable nonverbal feature group.
    \textit{vi)} Meeting environments, composed of 3-4 persons, were the most commonly investigated interaction scenario. Consequently, microphones and cameras were used more than other sensors to capture the data.
    \textit{vii)} The proposed methods, targeting the detection of the same phenomena, were rarely tested on common datasets.

Given the above observations, we further determine the following limitations.
    \textit{a)} Several datasets are not publicly available. This impairs the reproducibility of the results as well as not allows researchers to compare their method's performance with others.
    \textit{b)} There are datasets that lack annotation reliability analysis. Usage of them might result in misleading findings.
    \textit{c)} The datasets' scalability is not at the level of some other research domains using Artificial Intelligence. This can limit the application of advanced learning methods. Moreover, the models trained on limited material might not generalize well, thus they can perform well in some conditions but not in others.
    \textit{d)} The possible effectiveness of the deep models has not been investigated for several phenomena.
    \textit{e)} There were a limited number of attempts to perform cross-dataset analysis \cite{Aran2013b,Muller2019}, although there have been datasets supplying annotations for the same/related social phenomena, which were collected with similar sensor setups and scenarios allowing to extraction of the same nonverbal signals.
    Presenting such an analysis can give an idea regarding the potential performance of a proposed method during its real-life implementation as cross-domain analysis simulates the domain-shift problem.
    \textit{f)} Reviewed works lack explainability analysis and the usage of tools to improve interpretability.
    \textit{g)} We have not observed a discussion about how to transfer the presented human social interaction research into daily life by handling deployment time challenges such as domain-shift, and missing sensor data.

Motivated by the key observations and the limitations, we propose the future research directions:
      \textit{1)} adapting unsupervised feature pre-training, which can lower the need for labeled data compared to fully supervised methods,
    \textit{2)} integration of the domain adaptation methods for efficient data collection,
    \textit{3)} collecting more in-the-wild, long durational datasets, which would improve the generalizability of the developed models,
    \textit{4)} creating benchmark datasets to be used to compare different methodologies to extract and/or learn the nonverbal behaviors in a data-driven manner,
    \textit{5)} developing privacy-preserving nonverbal behavior representations, sensors, and computational models, and
   \textit{6)} implementing online learning methods and adaptive models in order to handle real-life challenges in a better way.
We believe that the research reviewed in this paper and the future work adapting our recommendations would result in effective Human Behavior Understanding that can be integrated into systems such as intelligent vehicles and social robots.

\appendix
\section{APPENDIX: Paper Selection Methodology}
\label{appendix1}
\begin{figure*}[!ht]
	\centering
	\includegraphics[width=0.8\linewidth]{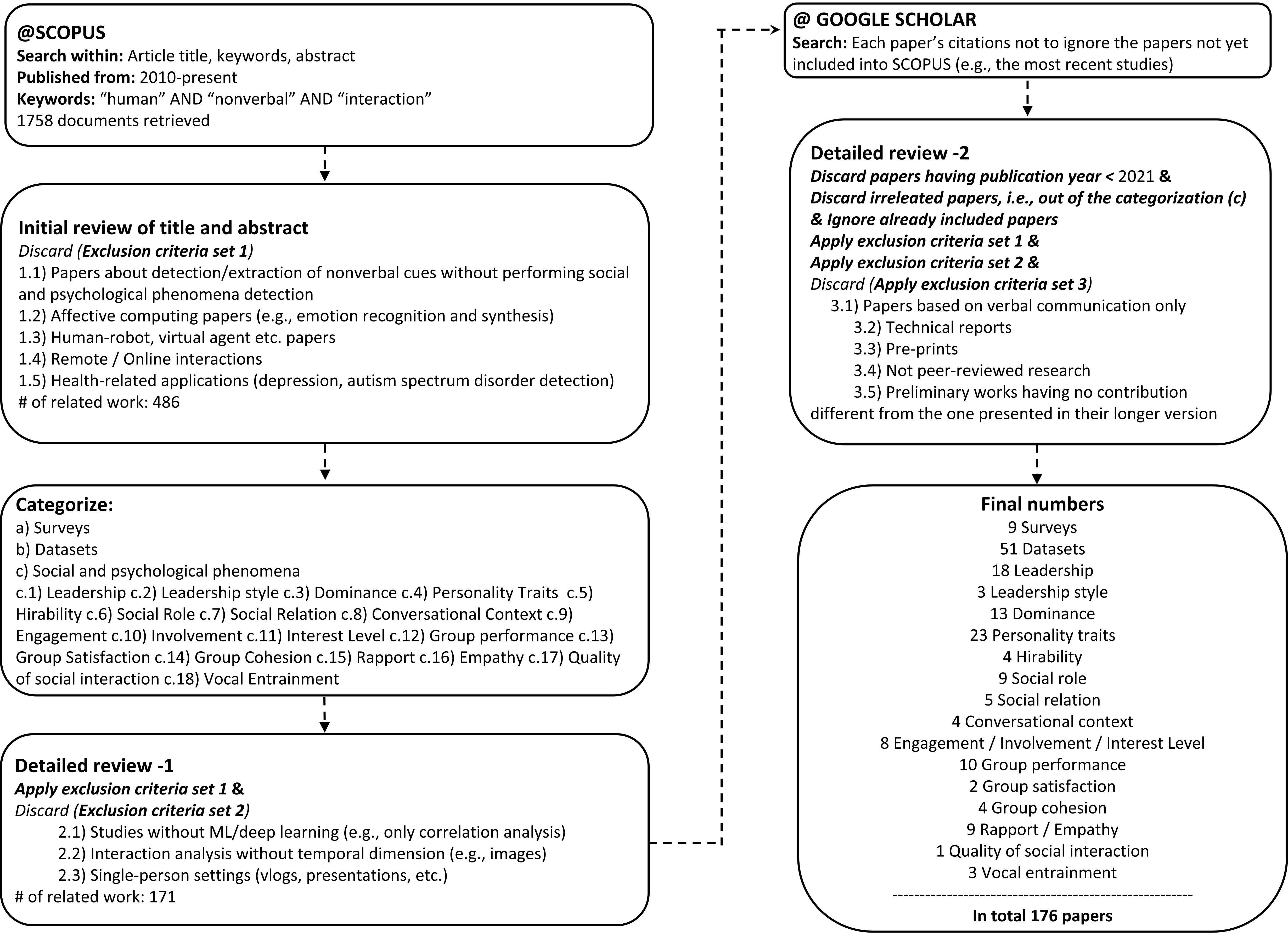}
	\caption{Selection process of the papers reviewed in this work.}
	\label{paperSelectFigure}
	\vspace{-0.5cm}
\end{figure*}

Fig. \ref{paperSelectFigure} demonstrates the paper selection methodology applied in this survey. We retrieved 1758 documents from SCOPUS with the search phrases ``human'' AND ``nonverbal'' AND ``interaction'' on titles, abstracts, and keywords when the date was selected higher or equal to 2010. Following that, we applied for an initial review of the titles and the abstract allowing us to discard the studies: a) detecting/extracting nonverbal cues without performing social and psychological phenomena, b) performing affective computing only, c) focusing on human-robot, virtual agents interactions, d) not applied on co-located interactions, e) focusing on health-related applications. As a consequence of this initial review, we obtained 486 documents. Such documents were categorized into surveys, datasets, and social and psychological phenomena methodology papers. We further made a detailed review in order to discard the papers such as not including a computational method (see Fig. \ref{paperSelectFigure} for more details). Meanwhile, we eliminated non-computing studies (e.g., psychology papers) as well. At that stage, we figured out that we missed some recent studies, perhaps because the proceedings they were involved in were not yet synchronized with SCOPUS. To confirm this, we additionally checked each paper's citations in GOOGLE SCHOLAR by limiting our review to the date "<" 2021. That allowed us to include 5 more peer-reviewed research papers matching our scope. Finally, our survey includes 116 computing studies in addition to 51 dataset papers. We also discuss 9 related surveys allowing us to express our contributions with respect to them.

\bibliographystyle{ACM-Reference-Format}
\bibliography{myBibliography}
\end{document}